\documentclass[%
 reprint,
nofootinbib,
 amsmath,amssymb,
 aps,
colorlinks=true, linkcolor=blue, citecolor=blue, urlcolor=blue,
]{revtex4-2}

\usepackage[utf8]{inputenc}
\usepackage{babel}
\usepackage{graphicx}
\usepackage{dcolumn}
\usepackage{bm}
\usepackage[mathlines]{lineno}
\usepackage{newtxtext}
\usepackage{newtxmath}
\usepackage{adjustbox}
\usepackage{subcaption}
\usepackage{amsfonts,amsmath, amssymb}
\usepackage{physics}
\usepackage{siunitx}
\usepackage{lipsum}
\usepackage{xcolor}
\usepackage{multirow}
\usepackage{pgfplots}
\usepackage{isotope}
\pgfplotsset{compat=1.18}
\usepackage{xspace}
\usepackage[T1]{fontenc} 
\usepackage{xparse}
\usepackage[normalem]{ulem}

\usepackage{caption}
\usepackage{newfloat}

\usepackage{algorithm2e}
\SetKwComment{Comment}{/* }{ */}
\SetAlgoLined
\SetAlgoNlRelativeSize{0}  
\RestyleAlgo{ruled}
\SetKwInput{KwInput}{Input}

\newcommand{\nuwro}{\textsc{NuWro}\xspace}
\newcommand{\neut}{\textsc{NEUT}\xspace}
\newcommand{\genie}{\textsc{GENIE}\xspace}
\newcommand{\gibuu}{\textsc{GiBUU}\xspace}
\newcommand{\ttok}{T2K\xspace}
\newcommand{\minerva}{MINER\ensuremath{\nu}A\xspace}
\newcommand{\microboone}{MicroBooNE\xspace}

\newcommand{\kamland}{KamLAND\xspace}
\newcommand{\sqjsns}{$\text{JSNS}^2$\xspace}

\newcommand{\exclusivedata}{CC1$p$0$\pi$\xspace}

\newcommand{\qe}{QE\xspace}
\newcommand{\mec}{MEC\xspace}

\newcommand{\spp}{SPP\xspace}
\newcommand{\dis}{DIS\xspace}
\newcommand{\hyp}{HYP\xspace}
\newcommand{\fsi}{FSI\xspace}

\usepackage{orcidlink}

\usepackage{hyperref}

\definecolor{JanRed}{rgb}{0.8, 0.0, 0.0}

\definecolor{HemantBlue}{rgb}{0.0, 0.2, 0.7}

\begin{document}

\preprint{APS/123-QED}

\title{Fine-tuning final state interactions model in \nuwro Monte Carlo event generator}


\author{Hemant Prasad \orcidlink{0009-0003-1897-9616}}
\email{hemant.prasad@uwr.edu.pl}

\author{Jan T. Sobczyk \orcidlink{0000-0003-4991-2919}}
\email{jan.sobczyk@uwr.edu.pl}


\author{Rwik Dharmapal Banerjee \orcidlink{0000-0003-3639-7532}}

\author{J. Luis Bonilla \orcidlink{0009-0009-3240-1494} }

\author{Krzysztof M. Graczyk \orcidlink{0000-0002-0038-6340}}

\author{Beata E. Kowal \orcidlink{0000-0003-3646-1653}}

\affiliation{   Institute of Theoretical Physics,
                University of Wroc{\l}aw, 
                Plac Maxa Borna 9, 50-204, 
                Wroc{\l}aw, Poland
            }

\author{Artur M. Ankowski \orcidlink{0000-0003-4073-8686}}
        \affiliation{ M. Smoluchowski Institute of Physics, Faculty of Physics, Astronomy, and Applied Computer Science, Jagiellonian University, Ul. Profesora Stanis{\l}awa {\L}ojasiewicza 11, 30-348 Krak{\'o}w, Poland}

\date{\today}

\begin{abstract}
Recent experimental data from \minerva on transverse kinematics observables across four different nuclear targets\textemdash carbon, oxygen, iron, and lead\textemdash have been utilized to refine the modeling of final state interaction effects in the \nuwro Monte Carlo neutrino event generator. For this purpose, we have developed an event reweighting tool for future applications to adjust the strength of final-state interactions. This study highlights the requirement for stronger nucleon reinteractions than previously assumed, but it still falls within the uncertainty range observed in a study comparing proton transparency measurements. This conclusion has significant implications for both experimental and theoretical work involving \nuwro.
\end{abstract}

\maketitle

\section{\label{sec:Introduction}INTRODUCTION}

\par 
Reducing uncertainties in neutrino–nucleus cross sections~\cite{ALVAREZRUSO20181,Katori_2018} is essential for the precision goals of forthcoming oscillation experiments~\cite{Abi_2020, abi2020deepundergroundneutrinoexperiment, eurJ-Abe2023}. With the growing event statistics, neutrino cross sections have already become the dominant source of uncertainties in current experiments~\cite{PhysRevD.106.032004, PhysRevLett.130.011801}. These experiments rely heavily on Monte Carlo (MC) event generators ~\cite{ParticleDataGroup:2024cfk} to simulate neutrino–nucleus interactions and to estimate efficiencies, backgrounds, and systematic uncertainties in experimental results. \\

\par 
The neutrino MC event generators that are widely used, such as \genie~\cite{ANDREOPOULOS201087}, \neut~\cite{Hayato}, \gibuu~\cite{BUSS20121}, or \nuwro~\cite{Golan:2012wx}, are based on the factorization scheme. Neutrinos interact with individual bound nucleons, and the overall interaction is separated into two sub-processes. The first is the primary (anti)neutrino-nucleon interaction, and the second is the rescattering of hadrons produced at the primary vertex. During \fsi, where the outgoing hadrons re-interact in various ways with the residual nucleus, they alter the event topology. For instance, a two-nucleon knockout could be due to either a \textit{2p-2h} interaction or by a quasielastic interaction followed by re-scattering of the struck nucleon, or maybe even from a pion-production mechanism followed by pion-absorption.\\

\par
Metropolis \textit{et al.}~\cite{PhysRev.110.185, PhysRev.110.204}, building on an idea of Serber~\cite{PhysRev.72.1114}, attempted in the early 1950s to describe these processes using what was then a novel MC approach within a cascade model of hadron propagation in the nuclear medium. Subsequently, many significant advancements were made in this direction, rendering the approach more realistic. Within the cascade model, the most critical component is the \textit{microscopic} hadron–nucleon cross section. Interacting hadrons experience a nuclear potential, so the free hadron–nucleon cross sections used in the original Metropolis \textit{et al.} papers were modified accordingly. Several approaches exist to model these microscopic cross sections, see e.g. Refs~\cite{PhysRevC.45.791, Li:1993rwa}, but the ultimate test of their effectiveness within a given framework must come from experiments.\\

\par 
In this paper, we study the \nuwro \fsi model. \nuwro is a widely used MC tool employed in many experimental and comparative studies. We will focus exclusively on the nucleon-\fsi part, deferring improvements in pion and hyperon modeling to future studies. Any improvements to neutrino interaction modeling in \nuwro can also be applied to other MC generators. We aim to provide a more comprehensive description of the overall strength of nucleon-\fsi effects in the \nuwro MC generator. In the cascade model, nucleons moving through the nucleus usually undergo only a few interactions. Accurately estimating their interaction probabilities is crucial. Other details, such as the angular distribution of outgoing nucleons after re-interaction, are less significant. The number of reinteractions, proportional to interaction probability, influences the kinetic energy of the nucleons, which is the most important observable. \\

\par In the previous comprehensive \nuwro nucleon-\fsi study, a comparison was done to proton transparency data~\cite{PhysRevC.100.015505}. The outcome of this study was an estimate of the overall uncertainty in quantifying the strength of the FSI effects, expressed as uncertainty in the value of the nucleon mean-free path. In conclusion, the estimate was $\pm 30\%$.  A recent study by  W. Filali \textit{et al.} ~\cite{PhysRevD.111.032009} on comparison of neutrino interaction models (and their implementation in different MC generators) with measurements of \textit{transverse kinematic imbalance} variables \cite{PhysRevD.92.051302, Lu_2016} from the \ttok, \microboone and \minerva experiments suggest that the \fsi models in MCs should be stronger, compared to their current status. In this paper, we examine information contained in the recent \minerva experimental measurements of transverse kinematic imbalance (TKI) observables ~\cite{minerva-paper}, in which they performed simultaneous measurements on four different-sized nuclei: carbon, oxygen, iron, and lead. In heavier nuclei, nucleons travel longer distances, therefore are more likely to reinteract, making them more sensitive to \fsi. For a nucleon with momentum $\sim 0.6$ GeV/c, the transparency ranges from 0.77 for carbon to 0.61 for iron and 0.42 for lead.\\

\par Recent \minerva data from \cite{minerva-paper} discussed in this article is expected to provide valuable information on the strength of \fsi effects. The main result of this paper is an estimate of the strength of \fsi in NuWro, which is $\sim 24\%$ larger than previously assumed, but remains consistent with the previously determined uncertainty.\\

To achieve our goal, we implement a reweighting scheme that allows us to modify the cascade's effective strength without requiring costly full-event resampling. The scheme reweighs only the nucleon reinteraction part, i.e., nucleon-\fsi. We reweigh the scaling parameter, which modifies the mean free path, thereby affecting the probability of nucleon-nucleon reinteraction. We explore different options for the amount of information per event that must be retained to enable reweighting. Using the proposed method, we then fine-tune \nuwro's cascade module to the recent \minerva's \exclusivedata data~\cite{minerva-paper}. \\

\par 
The paper is organized as follows. In Section~\ref{sec:nuwro}, we describe basic components of the \nuwro MC generator, focusing on its \fsi module (see Sect.~\ref{subsec:nuwro_cascade_model}). In Section~\ref{sec:minerva_data}, details about the recent \minerva \exclusivedata measurement ~\cite{minerva-paper} used in this study are presented. A separate Section~\ref{sec:reweighting-methods} is devoted to introducing the idea of reweighting in MC generators.  
In Section~\ref{sec:Results} we present the main results of this paper, how the \minerva data can be used to determine the strength of \fsi effects. The paper finalizes with discussion in Section \ref{sec:Discussion} and conclusions in Section \ref{sec:Conclusion}.\\

\section{\label{sec:nuwro}\nuwro MC EVENT GENERATOR}

\subsection{\label{subsec:nuwro-generalities} Generalities}

\par 
\nuwro~\cite{Juszczak:2005zs} is a MC event generator for lepton–nucleus interactions developed at the University of Wroc{\l}aw since $\sim$~2005 for accelerator-based neutrino experiments. It covers a broad energy range, from a few hundred MeV up to several hundred GeV. For primary interactions, \nuwro simulates interaction mechanisms such as quasielastic (\qe), quasielastic hyperon production (\hyp), meson-exchange currents (\mec), resonance production (RES), and deep inelastic scattering (\dis) for both charged-current (CC) and neutral-current (NC) processes. In the \qe channel, for both electron and neutrino
scattering, depending on the nucleus, \nuwro\ offers several options to describe target-bound nucleon: hole spectral function~(SF)~\cite{BENHAR1989267, BENHAR1994493}, effective density and momentum dependent potential~\cite{Juszczak:2005wk}, global (GFG) and local Fermi gas (LFG). Recently, \nuwro has been utilized as the primary MC generator in two significant experimental studies. The first was the \kamland measurement of the strange axial form-factor~\cite{KamLAND:2022ptk}. The second was \sqjsns measurement of missing energy with KDAR monoenergetic neutrino flux~\cite{JSNS2:2024bzf}.
\\

\par In this study \nuwro simulations are done with the version~\verb+25.03+. Compared to earlier distributions, there have been significant improvements in the \qe~\cite{PhysRevD.109.073004}, MEC~\cite{PhysRevD.111.036032}, and in single-pion production (\spp) \cite{ghent_hybrid_model}. In the \qe channel, the grid SF for argon is updated. In \mec channel \nuwro offers the implementation of a model developed by the Valencia group \cite{PhysRevC.102.024601} which has \textit{np-nh} decomposition of the total \mec cross section, isospin decomposition, and kinematic predictions for the two-nucleon final state. For \spp, \nuwro now uses, by default, the Ghent hybrid model~\cite{GonzalezJimenez:2016qqq}, which provides realistic  description of the second resonance region, interference with the non-resonant background, and a Regge description at high invariant hadronic masses. For the analysis, we use the LFG model to represent the initial state of the nucleon in \nuwro, since spectral-function implementations are available only for carbon, oxygen, argon, and iron, not for lead.\\

\subsection{\label{subsec:nuwro_cascade_model} Nucleon-\fsi module in \nuwro}

\par The \nuwro's cascade model offers a semi-classical description of hadron propagation in the nuclear medium, including nuclear effects such as in-medium modification of the hadron-nucleon cross section, Pauli blocking, and nucleon-nucleon correlations. Its basic framework is based on the MC approach proposed by Metropolis \textit{et al.}~\cite{PhysRev.110.185,PhysRev.110.204}. The model assumes that the energies transferred during collisions are much larger than the typical nuclear binding energy, allowing hadrons to be treated as quasi-free particles. The wave functions of the propagating hadrons are well-defined by their positions and momenta, and their de Broglie wavelengths are smaller than the mean free path. Additionally, scatterings on different nucleons are considered independent processes, with no interference between them. In Ref.~\cite{Yariv2007}, it has been shown that, for nucleons with sufficiently high momentum, these assumptions hold.\\

\par In \nuwro, the nuclear radius is defined as the distance from the center to the points where the density falls below the maximum density by a factor of $10^{5}$. The density profiles of the nuclei are sourced from Ref.~\cite{DEVRIES1987495}. The total and elastic free nucleon–nucleon cross sections are fitted to experimental data from Tanabashi 	\textit{et al.}~\cite{PhysRevD.98.030001}, while the fraction of single-pion production within the inelastic cross section follows the parametrization proposed by Bystricky \textit{et al.}~\cite{JPhys.France.Bystricky}. In-medium modifications to nucleon–nucleon interactions utilize the studies of Pandharipande \textit{et al.}~\cite{PhysRevC.45.791}, incorporating corrections for Pauli blocking and the effective mass of nucleons. For inelastic interactions, a density-dependent phenomenological correction~\cite{PhysRevC.48.1982} to the in-medium cross section ($\sigma_{\mathrm{NN}}^{*}$) is introduced as:
\begin{equation}
    \label{eqn:in-elastic-NN-crossection}
    \sigma_{\mathrm{NN}}^{*} (r)= \left(1 - \eta\,\frac{\rho (r)}{\rho_0}\right) \, \sigma_{\mathrm{NN}}^{\mathrm{free}}
\end{equation}
where $\rho_{0}$ is the saturation nuclear density and $\eta=0.2$.\\

\subsubsection{\label{subsubsec:nucleon-fsi-algorithm} Algorithm}

\par In \nuwro, the distance $\Delta x$ traveled by a propagating hadron between two successive possible interaction points is limited by a fixed step size of $0.2\,\text{fm}$. Over such a short distance, the nuclear density can be assumed to remain approximately constant. Under this assumption, the probability that a hadron travels a distance $\Delta x$ without undergoing an interaction, hereafter referred to as the \textit{survival probability}, is given by
\begin{equation}
\label{eqn:interaction-probability}
\tilde{\text{p}}(\Delta x) = \exp(-\Delta x / \lambda),
\end{equation}
where $\lambda = (\rho \sigma)^{-1}$ denotes the mean free path, computed locally from the nuclear density $\rho$ and the microscopic hadron--nucleon cross section $\sigma$. We reserve the symbol `p' (without tilde) for the interaction probability, defined as $\text{p} = 1 - \tilde{\text{p}}$. Reducing the step size further does not alter the physics results, but it significantly increases the computational runtime.

\par One common approach is to treat the actual travel distance as a random variable determined by the mean free path. In this formulation, $\Delta x$ is drawn according to
\begin{equation}
\label{eqn:free-path}
\Delta x = \min(-\lambda \ln{\xi}, \text{step}), 
\: \text{where} \: \xi \in (0, 1],
\end{equation}
with $\xi$ a uniformly distributed random number. If the sampled value satisfies $\Delta x < \text{step}$, an interaction is generated at that point; otherwise, the hadron is propagated freely to the next position a distance `step' away.

\par Alternatively, one may choose to fix $\Delta x = \text{step}$ deterministically. In this case, the decision to generate an interaction at the end of the step is taken by comparing the survival probability with the random number $\xi$: an interaction occurs if and only if $\tilde{\text{p}} < \xi$.

Both approaches lead to practically identical results. Keeping in mind our plan to develop the \fsi reweighting tool, in this study, we adopt the second approach where we fix $\Delta x=\text{step}$.\\

\subsubsection{ Proton Transparency Constraints on nucleon-\fsi} 

\par To quantify the distortion induced by \fsi effects, one can use the concept of \textit{nuclear transparency}, defined as the probability that a struck nucleon escapes the nucleus without undergoing significant reinteractions. Theoretical predictions are compared with experimental measurements by evaluating the MC transparency, defined as the fraction of simulated events in which the outgoing nucleon undergoes no reinteractions. It is important to note, however, that MC transparency is not directly equivalent to nuclear transparency, since experimental measurements cannot distinguish events with no reinteractions from those involving only ``soft’’ FSI; see the discussion in Ref.~\cite{PhysRevC.100.015505}. A detailed study of the uncertainties associated with modeling nucleon-\fsi within the \nuwro cascade framework was done under the assumption that the dominant source of uncertainty in the description of nucleon-\fsi could be effectively parameterized by scaling the nucleon mean free path. It was estimated that variations of approximately $\pm 30\%$ (see Fig.3 of Ref.~\cite{PhysRevC.100.015505}) covered the experimental uncertainty in the transparency measurements. To incorporate this uncertainty within \nuwro, a free parameter was introduced, denoted by $s$, which linearly scales the nucleon mean free path as $\lambda \rightarrow \lambda’ = s\lambda$ with $s \in [0.7, 1.3]$. 
A smaller (larger) value of $s$ corresponded to a shorter (longer) mean free path, leading to a higher (lower) interaction probability within the cascade and thus strengthening (weakening) the overall impact of \nuwro’s nucleon-\fsi. \\

\section{\label{sec:minerva_data} \minerva \exclusivedata DATA}

\par
Recently, the \minerva Collaboration published results based on the simultaneous measurement of the $\nu_{\mu}$ quasielastic-like (\exclusivedata) cross section across
different nuclear target materials—carbon, oxygen, iron, and lead~\cite{minerva-paper}. These results are obtained from events in which both a muon and a proton are reconstructed in the final state. The \minerva study uses the neutrinos-at-main-injector (NuMI) beam  in its medium-energy configuration~\cite{PhysRevD.107.012001}, characterized by average neutrino energy $\langle E_{\nu}\rangle \sim 6$ GeV with a spread of $\sim 2$ GeV. \\ 

\par
A detailed description of the \minerva detector appears in Ref.~\cite{ALIAGA2014130}. An important feature for this analysis is the radial variation of the neutrino flux within the detector, which varies by a few percent. This variation arises because the detector’s symmetry axis is not perfectly aligned with the beam axis, resulting in longitudinal variations in flux across the target. As a result, the flux differs for each nuclear target. The variation is most pronounced around $8\,\text{GeV}$, slightly away from the peak of the neutrino flux. The \minerva analysis accounts for these variations, and the published data incorporate this information. For consistency, we follow the same procedure in our \nuwro simulations by using target-dependent fluxes. \\

\par 
The \minerva study~\cite{minerva-paper} focuses on \qe-like processes of the form
\begin{equation}
\label{eqn:reaction}
\nu + A \to \mu^{-} + p + X,
\end{equation}
where $A$ is the target nucleus and $X$ is the final-state hadronic system, consisting of the remnant nucleus with possible additional protons or neutrons but with no mesons or heavy baryons. The signal definition selects events with no pions, one muon, and at least one proton, subject to constraints
\begin{eqnarray}
    2.0 < |\mathbf{p}_{\mu}|  < 20\, (\text{GeV}/c) \quad\quad \theta_{\mu} < 17^\circ, \label{eqn:muon-cut} \\
    0.5 < |\mathbf{p}_{p}| < 1.1\, (\text{GeV}/c)  \quad\quad \theta_{p} < 70^\circ,
    \label{eqn:proton-cut}
\end{eqnarray}
where $\mathbf{p}_{\mu}$ and $\theta_{\mu}$ ($\mathbf{p}_{p}$ and $\theta_{p}$) denote the outgoing muon (proton) momentum and polar angle relative to the incoming neutrino direction. For events with multiple protons satisfying the selection, the proton with the highest momentum is chosen. \\ 

\par 
Proton kinematics provide information that enables reconstruction of the transverse momentum imbalance relative to the neutrino direction. The projection of the momentum difference between the outgoing lepton and proton in this plane reflects nuclear effects that modify the proton’s kinematics. These quantities—known as TKI variables~\cite{Lu_2016, PhysRevC.95.065501, PhysRevD.92.051302}—are widely used in experimental analyses to characterize observed nuclear effects. For the process in Eq.~(\ref{eqn:reaction}), where the incident $E_{\nu}$ is unknown, the transverse momentum imbalance is
\begin{equation}
    \delta\mathbf{p}_{T} = \mathbf{p}^{\mu}_{T} + \mathbf{p}^{p}_{T},
\end{equation}
where $\mathbf{p}^{\mu(p)}_{T}$ is the transverse projection of the outgoing muon (proton) momentum. The imbalance can arise from Fermi motion or FSI. One can further decompose $\delta\mathbf{p}_{T}$ into x- and y-components. The standard convention in the neutrino community is to take the z-axis along the direction of the incoming neutrino and define

\begin{eqnarray}
    \delta\text{p}_{T_{y}} & = \hat{y}\cdot\delta\hat{\mathbf{p}}_{T},\qquad \hat{y} = -\hat{\mathbf{p}}^{\mu}_{T}, \\
    \delta\text{p}_{T_{x}} & = \hat{x}\cdot\delta\hat{\mathbf{p}}_{T},\qquad \hat{x} = \hat{z}\cross\hat{y}. 
\end{eqnarray}

Another important observable is the direction of $\delta\mathbf{p}_{T}$ relative to the transverse projection of the muon momentum ($-\hat{\mathbf{p}}_{T}^{\mu}$), denoted $\delta\alpha_{T}$:

\begin{equation}
    \delta\alpha_{T} = \arccos{\left(-\hat{\mathbf{p}}^{\mu}_{T}\cdot\delta\hat{\mathbf{p}}_{T}\right)}.
\end{equation}

A further TKI variable commonly used in experimental studies is $\delta\phi_{T}$, which measures the direction of the outgoing nucleon relative to the momentum-transfer vector $\vec{q}$ in the transverse plane:

\begin{equation}
    \delta\phi_{T} = \arccos{\left(-\hat{\mathbf{p}}^{\mu}_{T} \cdot \hat{\mathbf{p}}^{p}_{T}\right)}.
\end{equation}

\par One can also go beyond TKI and explore information contained in longitudinal components of proton and muon, to reconstruct the momentum of the struck neutron \cite{PhysRevC.95.065501}. The estimate is very precise for \qe interactions without \fsi. Using the energy-momentum conservation, we get:

\begin{equation}
|\mathbf{p}_{n}| = \sqrt{|\delta\mathbf{p}_{T}|^2 + |\delta\mathbf{p}_{L}|^2}
\end{equation}
where 
\begin{equation}
    |\delta\mathbf{p}_{L}| = \displaystyle \frac{R^{2} - m_{A'}^{2} - \delta\mathbf{p}_{T}^{2} }{2R}
\end{equation}
and
\[R=m_{A} + |\mathbf{p}_{L}^{\mu}| + |\mathbf{p}_{L}^{p}| - E^{\mu} - E^{p}\,.\]
$m_{A^{(')}}$ and $E^{\mu(p)}$ are the mass of the nuclear target (remnant nucleus),
and the total energy of the muon (proton). In the \minerva study, 

\[m_{A'} = m_{A} - m_{n} + b\]

where $m_{n}$ is the mass of the neutron and $b$ is the excitation energy with the following values: 27.13 MeV for carbon, 24.1 MeV for oxygen, 29.6 MeV for iron, and 22.8 MeV for lead, obtained from the Ref.~\cite{PhysRevC.95.065501}. Section~\ref{subsec:Nuwro_vs_data} provides a physical interpretation of the TKI variables used in our analysis when comparing \nuwro predictions with the \minerva's \exclusivedata data.\\

\par 
The \minerva data release includes measurements of differential cross sections for TKI observables and for the observed muon and proton kinematics, including momenta, transverse and longitudinal components, and angular distributions relative to the beam, as well as ratios of differential cross sections to the hydrocarbon target. Each observable is accompanied by an overall covariance matrix, which is used to compute the $\chi^{2}$ per degree of freedom (d.o.f) in comparisons with MC predictions. The collaboration also provides a covariance matrix combined over all nuclear targets for each observable.

\section{\label{sec:reweighting-methods}REWEIGHTING OF \fsi}

\par
Any change in the \fsi parameters alters the likelihood of events produced by an MC generator. Instead of rerunning the MC with a new configuration, we can achieve the same effect by directly modifying the probability (weight) of each event. This process, known as reweighting, is crucial when detector simulations are time-consuming and when we must estimate how parameter changes affect observed experimental outcomes. For example, consider the reweighting procedure for modifying a theoretical parameter such as the \qe axial mass $M_A$ in a dipole parameterization. In this case, we compute the axial form factor by evaluating the ratio of differential cross sections for two different values of $M_A$ in each event, using the available information about the four-momentum transfer $Q^2$ and differential cross section $d\sigma/d Q^2$. However, reweighting \fsi parameters operates under different principles. We describe this process below. \fsi reweighting has been thoroughly studied before in the context of the \neut MC generator, see Ref.~\cite{Nonnenmacher_2020}.

\subsection{\label{subsec:exact-scheme} \textit{Exact}-reweighting}

\par
During the runtime of \nuwro's cascade module, the MC algorithm checks at each step whether any propagating nucleons interact. This decision is based on the survival probability $\tilde{\text{p}}$, which depends on the nuclear density and the effective nucleon–nucleon cross section.\\

\par
For \textit{exact} reweighting, we classify each cascade step into one of the following categories:
\begin{enumerate}
    \item A step where an interaction occurred.
    \item A step where an interaction would have occurred but was prevented due to Pauli blocking.
    \item A step where no interaction occurred.
\end{enumerate}
From the exact-reweighting viewpoint, we treat Pauli-blocked interactions the same way as regular interaction steps.\\

\par
We can view the \fsi part of an event simulation as a chain of interaction and non-interaction steps: 
\[ \tilde{\text{p}}_1, \tilde{\text{p}}_2, ..., \tilde{\text{p}}_j,  \text{p}_1, \tilde{\text{p}}_{j+1}, ...,\text{p}_{N_{s_i}}...,\tilde{\text{p}}_{N_{f_i}}\, .\]
These chains tend to be long because nucleons typically travel through many steps before leaving the nucleus. For example, simulations that use lead as the target can produce chains with $\sim 3000$ or more steps. A complete information about \fsi is contained in all $\text{p}_k$, and $\tilde{\text{p}}_j$. However, for a purpose of reweighting it is enough to store all $\tilde{\text{p}}_j$ as a single value of the product of $\tilde{\text{p}}_{j}$'s as:
\[ \tilde{\text{P}} = \prod_{l}\tilde{\text{p}}_{l}. \]  
The essence of FSI reweighting is to modify the interaction and non-interaction probabilities in a controlled manner so that the likelihood of a given event changes accordingly.\\

\par
We define the likelihood of the $i^{\mathrm{th}}$ event, observed with $N_{s_i}$ interaction steps and $N_{f_i}$ non-interaction steps, as:
\begin{equation}
\label{eqn:likelihood-exact}
    \mathcal{L}(N_{s_i}, N_{f_i}; \{\text{p}_k\}, \{\tilde{\text{p}}_l\}) 
    = \underbrace{\prod_{l=1}^{N_{f_i}} \tilde{\text{p}}_l}_{\tilde{\text{P}}} \,
      \prod_{k=1}^{N_{s_i}} \text{p}_k .
\end{equation}

\par
To reweight the strength of the \fsi, we scale the mean free path of the propagating nucleons by a factor of $s$, see Eq.~(\ref{eqn:interaction-probability}). This modifies the interaction and non-interaction probabilities as follows:

\begin{equation}
    \label{eqn:scaling-interaction-probability}
    \vcenter{\hbox{$\lambda \to s\lambda \implies$}} 
    \quad 
    \begin{aligned}
        & \tilde{\text{p}} \to \tilde{\text{p}}' = (\tilde{\text{p}})^{1/s}, && \\
        & \text{p} \to \text{p}' = 1 - (1-\text{p})^{1/s} && 
    \end{aligned}
\end{equation}

A larger mean free path results in a lower interaction probability, while a shorter mean free path increases the interaction probability.

\par
The modified likelihood becomes
\begin{equation}
\label{eqn:likelihood-exact-prime}
    \mathcal{L}'(N_{s_i}, N_{f_i}; \{\text{p}'_k\}, \{\tilde{\text{p}}'_l\})
    = \underbrace{\prod_{l=1}^{N_{f_i}} \tilde{\text{p}}'_l }_{\tilde{\text{P}}'= \tilde{\text{P}}^{1/s}}\,
      \prod_{k=1}^{N_{s_i}} \text{p}'_k .
\end{equation}

\par
The reweighting factor for the event is therefore
\begin{equation}
\label{eqn:reweight_value}
    w'_i(s) = \frac{\mathcal{L}'}{\mathcal{L}} .
\end{equation}
Algorithm~\ref{alg:exact_reweighting_scheme} presents a complete schematic for reweighting \fsi.\\

\begin{algorithm}[htbp]
\caption{\textit{Exact}-reweighting scheme.}
\label{alg:exact_reweighting_scheme}

\KwInput{\nuwro\ output \texttt{.root} file with $M$ events.}

Choose the scale parameter $s$.

\For{event $i = 1$ \KwTo $M$}{
  \For{vertex $j = 1$ \KwTo $N$}{
    \eIf{interaction or Pauli-blocked vertex}{
      $\text{p}_j \rightarrow \text{p}'_j = 1 - (1 - \text{p}_j)^{1/s}$
    }{
      $\tilde{\text{p}}_j \rightarrow \tilde{\text{p}}'_j = \tilde{\text{p}}_j^{1/s}$
    }
  }
  Assign event with weight:
  $w'_i(s) = \mathcal{L}' / \mathcal{L}$\;
}

\KwResult{New \texttt{.root} file with updated event weights.}

\end{algorithm}

\par
We use two conditions to assess the consistency of the reweighting scheme:
\begin{itemize}
    \item \textbf{Condition 1:} Only the relative event weights should change. The average reweighting factor should satisfy
    \begin{equation}
        \frac{1}{M}\sum_{i=1}^{M} w'_i(s) \approx 1 .
    \end{equation}
    \item \textbf{Condition 2:} The outgoing lepton must remain unaffected. Therefore, differential cross sections expressed in lepton kinematic variables must not change.
\end{itemize}

\par
Both conditions must hold for the reweighting scheme to be reliable. Condition 2 is more difficult to check because it requires a sufficiently large number of events per bin to control statistical fluctuations. We examine this issue shortly in Sect.~\ref{subsec:scheme-consistency}.\\

\par To enable exact reweighting, the sequence of interaction probabilities and the product of survival probabilities must be recorded for every event. Storing this information significantly increases the size of the output file. We checked that including these data increases the file size by $60\%$. To address this issue, we will discuss the introduction of an \textit{approximate} reweighting scheme in Sect.~\ref{subsec:approx-scheme} that relies solely on information already stored in the current \nuwro version, such as the number of interaction and non-interaction steps.\\

\section{\label{sec:Results}RESULTS} 

\par In this section, we discuss the performance of the exact reweighting defined in Sect.~\ref{subsec:exact-scheme} and show how it can be used to fine-tune the strength of nucleon \fsi using \minerva \exclusivedata data~\cite{minerva-paper}.

\begin{figure}[htbp!]
    \centering
    \begin{subfigure}[b]{\linewidth}
         \centering
         \includegraphics[width=\linewidth]{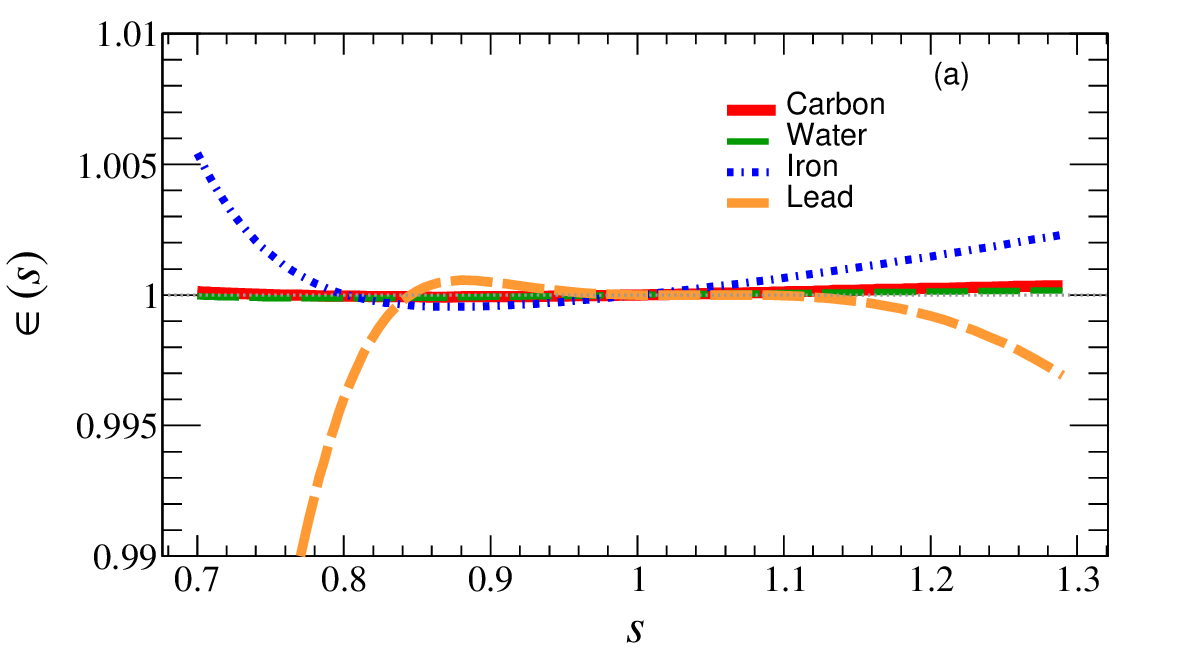}
     \end{subfigure}
     \hfill
     \begin{subfigure}[b]{\linewidth}
         \centering
         \includegraphics[width=\textwidth]{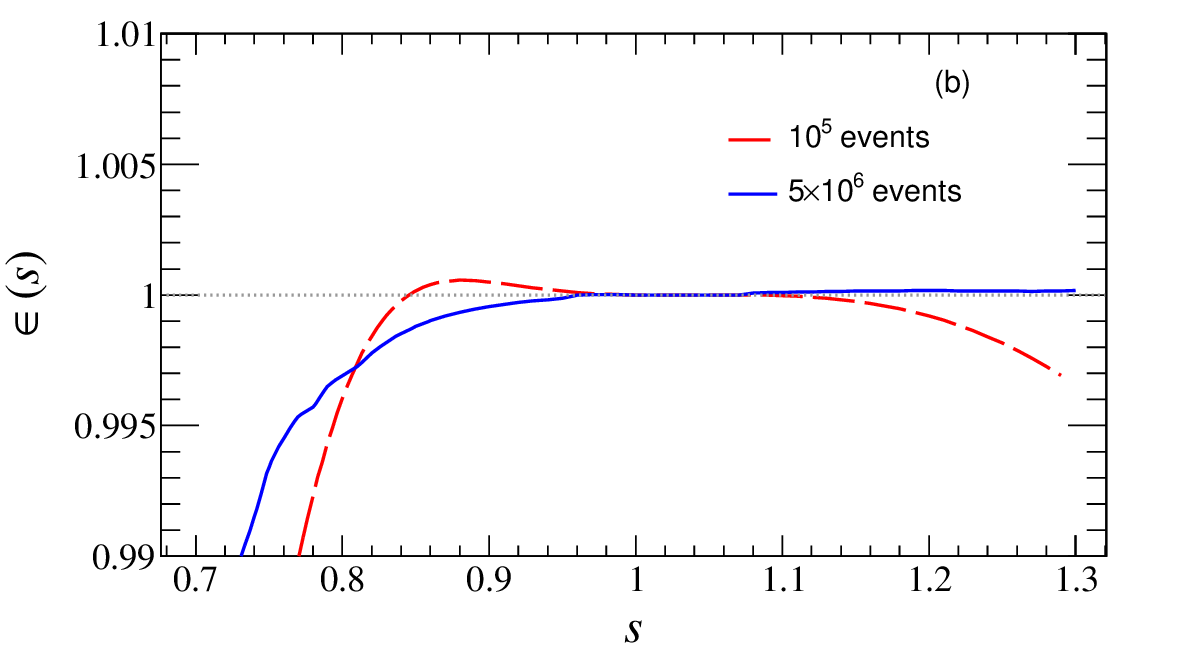}
     \end{subfigure}
    \caption{\label{fig:exact-scheme-performance} (Color online) Preservation of the overall normalization, as defined in Eq.~(\ref{eqn:scheme-performance}), for the method summarized in Algorithm~\ref{alg:exact_reweighting_scheme}. (\textbf{Top}) For different targets with $10^5$ events. (\textbf{Bottom}) For the lead target but with different statistics. }
\end{figure}

\subsection{\label{subsec:scheme-consistency} Consistency of the \textit{exact}-reweighting scheme}

\par We begin by defining a measure $\epsilon(s)$ to quantify how well the overall normalization is preserved under the reweighting scheme. It is given by

\begin{equation}
    \label{eqn:scheme-performance}
    \epsilon (s) = \frac{1}{M}\displaystyle\sum_{i=1}^{M} w'_{i}(s),
\end{equation}
where $M$ is the total number of events in the sample.\\

\par Figure~\ref{fig:exact-scheme-performance} shows the behavior of $\epsilon(s)$ for different nuclear targets. As seen in Fig.~\ref{fig:exact-scheme-performance}(a), the normalization is maintained within $1\%$ for all targets in a wide range of the parameter $s$. For carbon (red curve) and oxygen (green curve), the curves are almost flat with a minimal deviation. The normalization for iron (blue curve) is slightly overestimated, but remains within 1\% for the whole range of $s$. For lead (orange curve), the normalization remains stable in the interval $s \sim [0.79,\,1.3]$, but begins to deviate by more than $1\%$ for $s \lesssim 0.78$. This deterioration can be attributed to the limited statistics used for the reweighting. This is demonstrated in Fig.~\ref{fig:exact-scheme-performance}(b), where we show the effect of an increased number of events from $10^{5}$ to $5\times 10^{6}$. With the higher-statistics sample (blue curve in Fig.~\ref{fig:exact-scheme-performance}(b)), for $s>1$ the curve is flat, while for $s<1$ the range over which normalization is preserved up to 1\% expands to  $s \gtrsim 0.74$.\\

\par Similarly, we examined whether the differential cross section expressed in terms of lepton kinematic variables remains stable while reweighting within statistical fluctuations. To assess this, we compared the intervals in the scaling parameter $s$ over which the kinematic distributions remain shape-preserving across the two statistical samples considered above. Using a moderately fine binning, we checked whether distributions in variables such as $|\mathbf{p}_{\mu}|$ and $\theta_{\mu}$ agree to within $1\%$.\\

\par For the sample with $10^{5}$ events, the shape invariance condition is typically satisfied over the range $s \sim [0.82,\, 1.18]$, which is smaller than the interval in which normalization is retained. For the higher-statistics sample, the range over which the kinematic distributions remain stable broadens to $[0.78,\, 1.30]$. This demonstrates that the reweighting scheme's performance improves as the number of statistics increases. In the limit of infinite statistics, one expects the exact scheme to preserve both normalization and shape of kinematic distributions.\\

\par 
It is possible to overcome limitations caused by finite statistics in the reweighting range through an iterative process. To get results for smaller $s$ values, we run \nuwro with $s=0.76$ and reweight downward relative to this value, repeating the procedure if necessary.

\subsection{\label{subsec:fine-tune-nuwro} Fine-tuning \nuwro's cascade module}
\par To fine-tune \nuwro's cascade module, we applied the \textit{exact}-reweighting scheme (discussed in Sect.~\ref{subsec:exact-scheme}), which allows reweighting of FSI events without the need to rerun \nuwro. In this section, we describe the methodology used to optimize \nuwro's cascade module using the recent \minerva \exclusivedata data~\cite{minerva-paper}.

\subsubsection{\label{subsubsec:estimation-of-s} Finding the optimal value of `$s$'}

\par 
Our goal is to determine the value of the $A$-independent cascade parameter $s$. To achieve this, we compare \nuwro predictions across all four nuclear targets simultaneously. First, \nuwro output \texttt{.root} files are generated using the default cascade parameter value, $s = 1$, with the corresponding neutrino flux and the LFG nuclear model. We then select observables sensitive to \fsi modeling. Our choice of experimental observables is based on the following observations:
\begin{itemize}
    \item \fsi-reweighting only affects the kinematics of outgoing hadrons, keeping the lepton kinematics unaffected; therefore, observables based only on lepton kinematics cannot be used.
    \item Transverse projections such as TKI variables, as described in Ref.~\cite{Lu_2016}, and $|\mathbf{p}_{p_{T}}|$ provide direct measurements of nuclear effects and are included while optimizing $s$.
    
    \item More information about \fsi is contained in observables, which also involve the longitudinal components of final state particles, like reconstructed neutron momentum $|\mathbf{p}_{n}|$.
\end{itemize}

Based on these facts, we decided to select the following observables:

\begin{enumerate}
    \item TKI-variables $\delta\alpha_{T}, \delta\mathbf{p}_{T}, {\displaystyle {\delta\mathbf{p}_{T}}_{y}}, $ $\delta\phi_{T}$.
    \item Reconstructed neutron momentum $|\mathbf{p}_{n}|$ and transverse projection of the outgoing proton's momentum $|{\mathbf{p}_{p}}_{T}|$.
\end{enumerate} 

We then adopt the \textit{exact}-reweighting procedure to fine-tune the cascade parameter to these observables.\\


\begin{algorithm}[htbp]
\caption{Fine-tuning of $s$}
\label{alg:optimization_scheme}

\KwInput{\nuwro output \texttt{.root} file for all targets with $s = 1$}

\For{$s \in [0.3,\,1.3]$}{
  1. Reweight \nuwro events using the scheme given in
  Algorithm~\ref{alg:exact_reweighting_scheme}.

  2. Apply event selection cuts~\cite{minerva-paper}
  to reweighted events.

  3. Produce distributions of the desired observable and compute
  $\chi^{2}$/d.o.f.\ using the combined covariance matrix over all
  targets (see Sect.~\ref{sec:minerva_data}).
}

\end{algorithm}


\par To determine the optimal cascade parameter, we scan $s$ in a large interval $[0.3, 1.3]$ with a step of $\Delta s = 0.01$, calculating $\chi^2$ for all selected observables. In our analysis, we use the covariance matrix to compute the $\chi^{2}/\text{d.o.f}$ for each observable and to fine-tune \nuwro's cascade module with respect to the parameter $s$. For each observable, the fitting process follows the algorithm outlined in Algorithm~\ref{alg:optimization_scheme}. This procedure is then repeated for all observables considered. For each observable, we get a best fit value of $s$.\\

\par
\begin{figure}[htbp!]
    \centering
    \begin{subfigure}[b]{\columnwidth}
        \centering 
        \includegraphics[width=\linewidth]{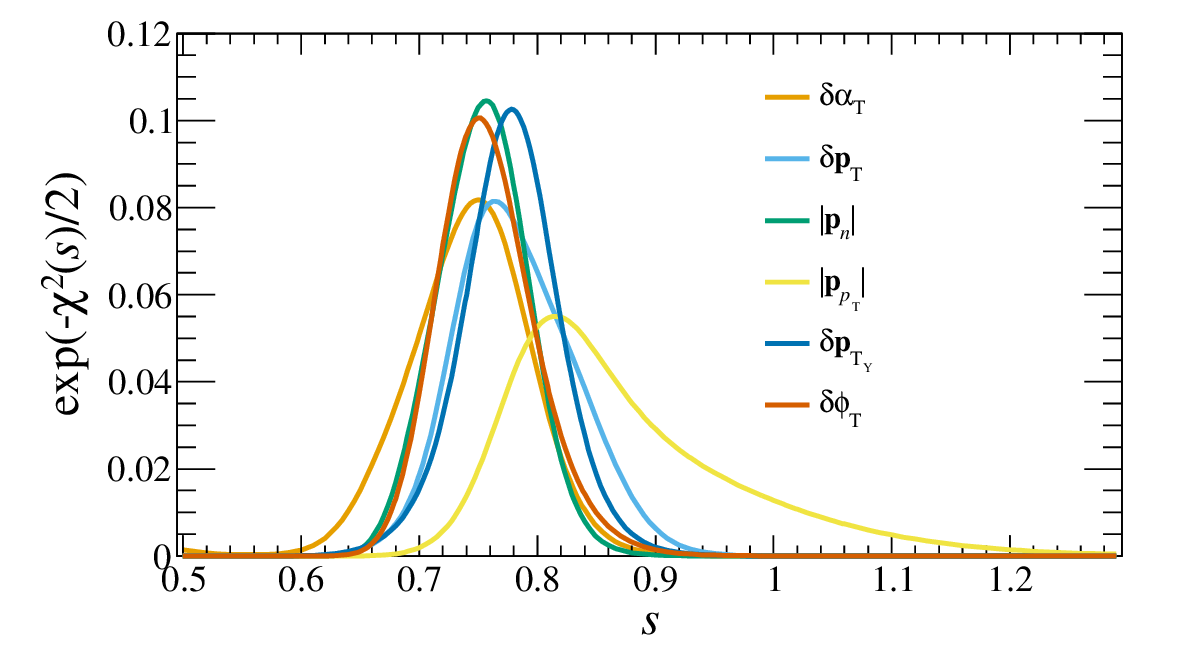}
    \end{subfigure}
    \hfill%
    \begin{subfigure}[t]{\columnwidth}
        \centering
        \includegraphics[width=\linewidth]{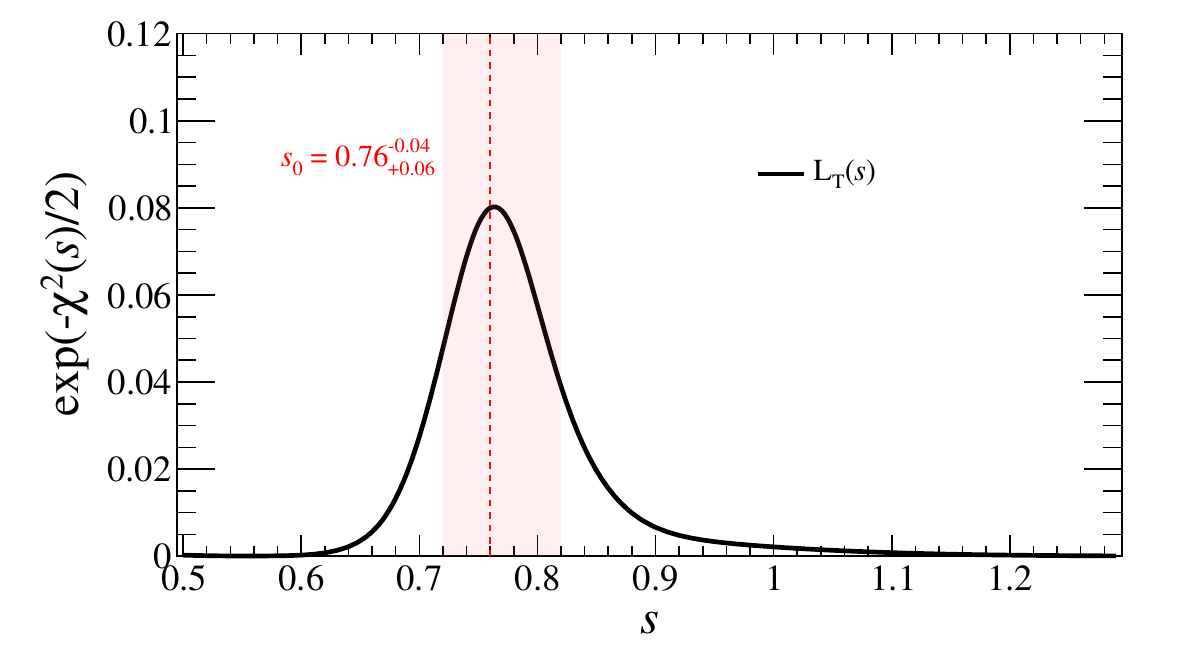}
    \end{subfigure}
    \caption{\label{fig:} (\textbf{Top}) Likelihood $\hat{L}(s)$ inferred using $\chi^2(s)$ for each observable. (\textbf{Bottom}) Overall likelihood $L_{T}(s)$ by summing over all observables. The red dashed line represents the peak. The red band represents the minimum length for which the enclosed area is $68.27\%$.}
    \label{fig:likelihood}
\end{figure}

 To estimate the best fit value of $s$ globally across all observables, along with its uncertainty, we examine the likelihood function inferred from the $\chi^{2}(s)$ functions associated with each observable. In statistics, the goodness of fit expressed through $\chi^{2}$ can be related to the likelihood ratio of two competing models. For a model with parameters \(\hat{\theta}\), having likelihood \(L(\hat{\theta})\), compared with a reference (null) model with likelihood \(L_{0}\), the likelihood ratio test gives  
\[
\chi^{2} \sim -2\ln{\left(\frac{L(\hat{\theta})}{L_{0}}\right)},
\]
Inverting the above relation yields a likelihood function \(L(\hat{\theta})\) for each observable based on its corresponding \(\chi^{2}\).  In our case, we compute the likelihood $L(s)$ of observing $s$ for a given observable, using the distribution of $\chi^{2}(s)$ for that observable. The likelihood distributions for different observables used in our analysis are shown in the top panel of Fig.~\ref{fig:likelihood}. By comparing results across six observables, we obtain a consistent picture: the parameter $s$ should be much smaller than its default value. Differences between observables are not surprising, keeping in mind the complexity of physics mechanisms giving rise to the experimental signal measured by the \minerva, and the fact that we focus on only one aspect of the \fsi model \textemdash the overall strength.\\

\par To obtain a global constraint on $s$, we define the overall likelihood $L_{T}$ as the (normalized) average of the individual likelihoods $L_{o}$ from all observables:
\begin{equation}
    L_{T} =\displaystyle \frac{1}{6}\cdot\sum_{o\in \left\{ \displaystyle\substack{\delta\alpha_{T}, \delta\mathbf{p}_{T},  {\delta\mathbf{p}_{T}}_{y}, \\ \delta\phi_{T},  |\mathbf{p}_{n}|, |{\mathbf{p}_{p}}_{T}| }\right\} } \hat{L}_{o},
\end{equation}
where the hat denotes that each likelihood has been normalized such that the area under the curve is unity. The resulting overall likelihood distribution is shown in the bottom panel of Fig.~\ref{fig:likelihood}. The peak of the likelihood is located at:

\begin{equation}
    s_{0} = 0.76
\end{equation}

The red band represents the $1\sigma$ uncertainty interval around $s_0=0.76$, defined as the section of minimal length such that the enclosed area is $68.27\%$ of the total, which is given as: 
\begin{equation}
    s \in [0.72, 0.82] \,.
\end{equation}
This will be the new uncertainty interval in the upcoming version of \nuwro. We emphasize that the new uncertainty interval is fully contained within the previous one, which was determined as $s\in [0.7, 1.3]$, as reported in Ref.~\cite{PhysRevC.100.015505}.

\subsection{\label{subsec:Nuwro_vs_data} Performance of \nuwro \fsi with $s=s_{0}$}

\begin{table*}[htbp!]
\caption{\label{tab:chi-square-table} $\chi^{2}$/d.o.f comparing \nuwro 
with \minerva's \exclusivedata data~\cite{minerva-paper} for different nuclear targets.}
\begin{ruledtabular}
\begin{tabular}{l | c c c c | c c c c}
\textrm{Observable} & \multicolumn{4}{c}{\nuwro ($s=1$)} & \multicolumn{4}{c}{\nuwro ($s = s_{0}$)} \\
\colrule
 & \textrm{Carbon} & \textrm{Water} & \textrm{Iron} & \textrm{Lead} & \textrm{Carbon} & \textrm{Water} & \textrm{Iron} & \textrm{Lead} \\
\colrule
1. $\delta\alpha_{T}$ 
& $14.29/5$ & $12.35/5$ & $19.11/5$ & $52.56/5$
& $13.05/5$ & $8.85/5$ & $8.89/5$ & $30.14/5$ \\
2. $\delta\mathbf{p}_{T}$ 
& $77.32/9$ & $21.91/9$ & $92.46/9$ & $53.02/9$
& $62.95/9$ & $19.82/9$ & $85.12/9$ & $37.16/9$ \\
3. $\delta\mathbf{p}_{T_{y}}$
& $16.80/9$ & $10.22/9$ & $20.11/9$ & $39.45/9$
& $14.82/9$ & $7.27/9$ & $10.45/9$ & $23.32/9$ \\
4. $|\mathbf{p}_{n}|$
& $59.46/10$ & $44.18/10$ & $56.88/10$ & $52.03/10$
& $50.23/10$ & $39.11/10$ & $54.23/10$ & $33.04/10$ \\
5. $\delta\phi_{T}$
& $39.56/6$ & $11.25/6$ & $25.03/6$ & $20.14/6$
& $29.11/6$ & $8.68/6$ & $14.11/6$ & $13.28/6$ \\
6. $|{\mathbf{p}_{p}}_{T}|$
& $5.22/7$ & $2.98/7$ & $13.48/7$ & $12.44/7$
& $4.52/10$ & $3.13/7$ & $9.39/7$ & $11.05/7$ \\
\hline
\multicolumn{9}{c}{\textrm{\textbf{Not used while optimizing $L_{T}(s)$}}}\\
\colrule
\textrm{Observable} & \multicolumn{4}{c}{\nuwro ($s=1$)} & \multicolumn{4}{c}{\nuwro ($s = s_{0}$)} \\
\colrule 
& \textrm{Carbon} & \textrm{Water} & \textrm{Iron} & \textrm{Lead} & \textrm{Carbon} & \textrm{Water} & \textrm{Iron} & \textrm{Lead} \\
\colrule
7. $|\mathbf{p}_{p}|$
& $12.75/5$ & $1.59/5$ & $19.62/5$ & $29.52/5$
& $12.18/5$ & $1.18/5$ & $15.16/5$ & $19.51/5$ \\
8. $\theta_{p}$
& $4.07/10$ & $5.55/10$ & $8.81/10$ & $21.94/10$
& $4.68/10$ & $5.80/10$ & $6.11/10$ & $17.19/10$ \\
\end{tabular}
\end{ruledtabular}
\end{table*}

\par \nuwro performance with $s=s_{0}$ for all the observables is summarized in Table~\ref{tab:chi-square-table}. The calculations were performed using individual covariance matrices for each target and each observable, separately. Table~\ref{tab:chi-square-table} presents information about six observables used in the fitting process, as well as two additional ones ($|\mathbf{p}_{p}|$) and $\theta_{p}$). The table supports our statement that the results are fully consistent. In nearly all cases, changing from $s=1$ to $s=s_{0}$ produces significantly better agreement with the data. The only exceptions occur when the $\chi^2$/d.o.f. ratio is below 1, which makes it problematic to draw firm conclusions.\\

\begin{figure*}[htbp!]
    \begin{subfigure}[b]{\columnwidth}
        \centering 
        \includegraphics[width=1.11\linewidth]{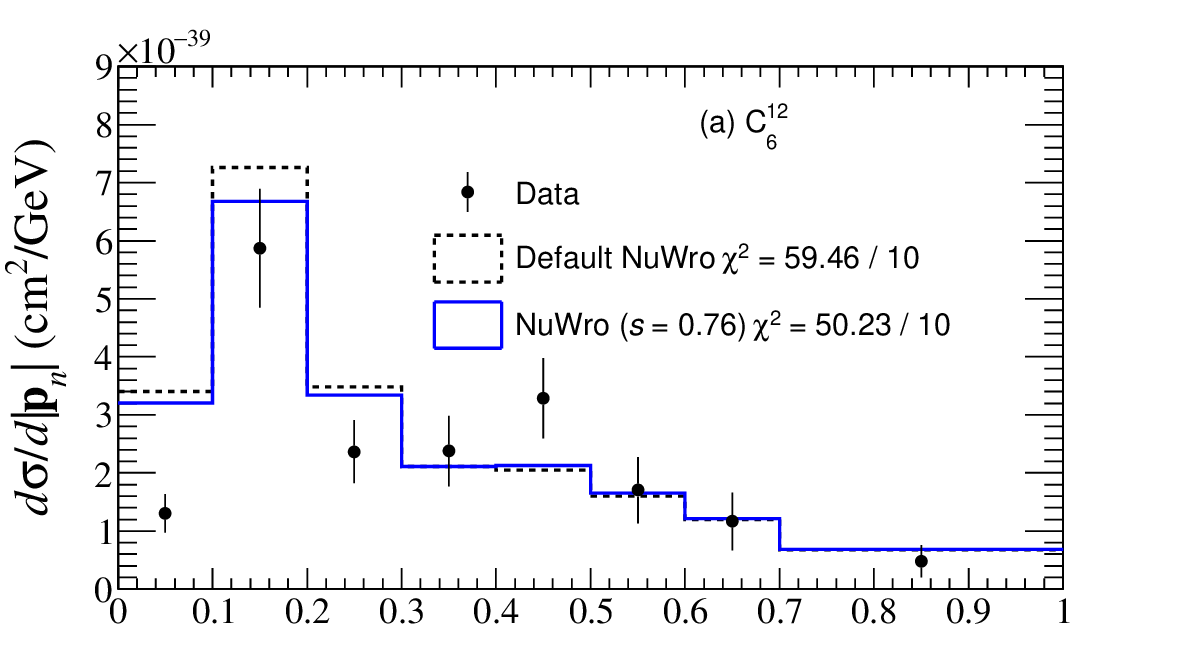}
    \end{subfigure}
    \begin{subfigure}[b]{\columnwidth}
        \centering 
        \includegraphics[width=1.11\linewidth]{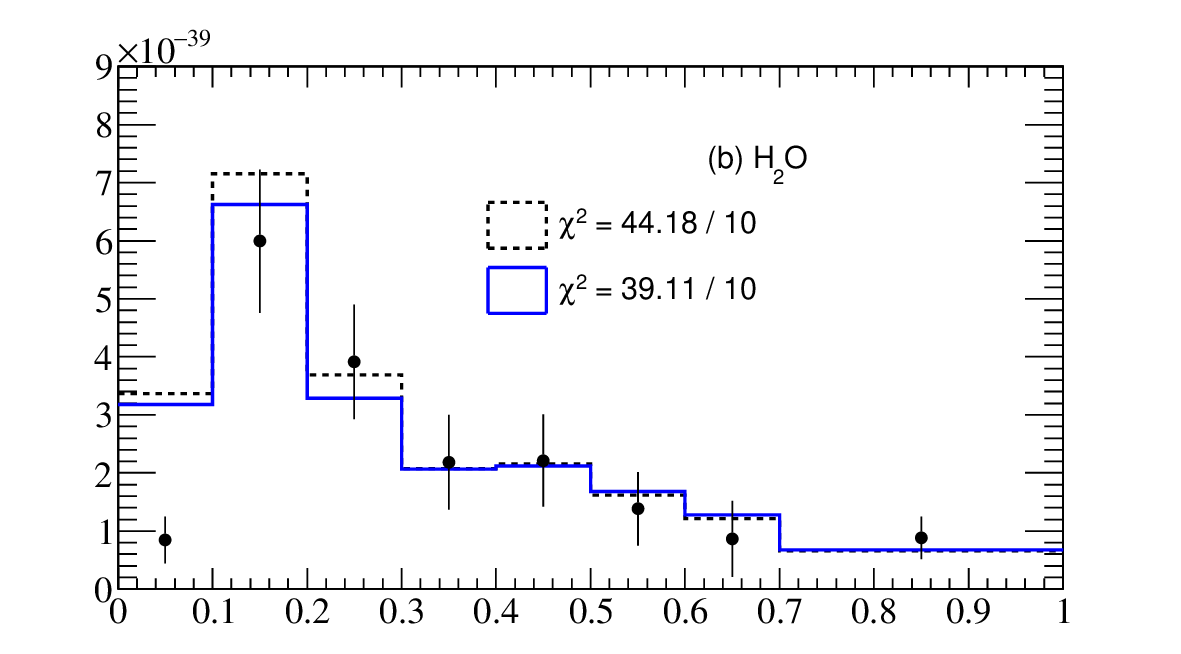}
    \end{subfigure}
    \begin{subfigure}[b]{\columnwidth}
        \centering 
        \includegraphics[width=1.11\linewidth]{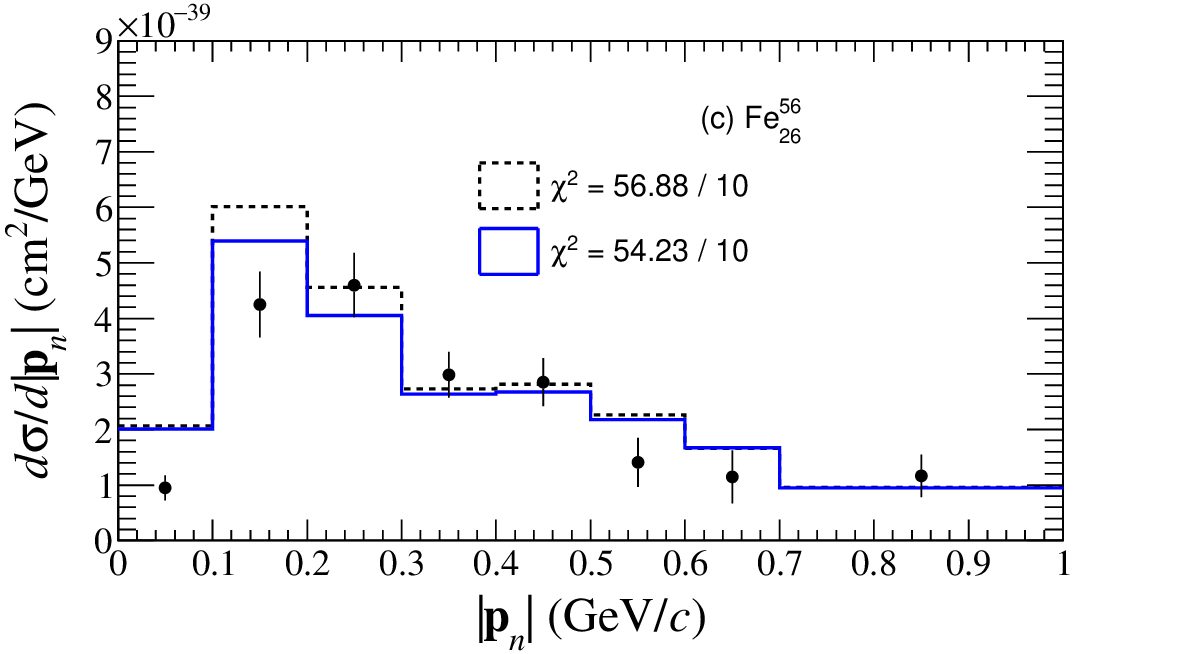}
    \end{subfigure}
    \begin{subfigure}[b]{\columnwidth}
        \centering 
        \includegraphics[width=1.11\linewidth]{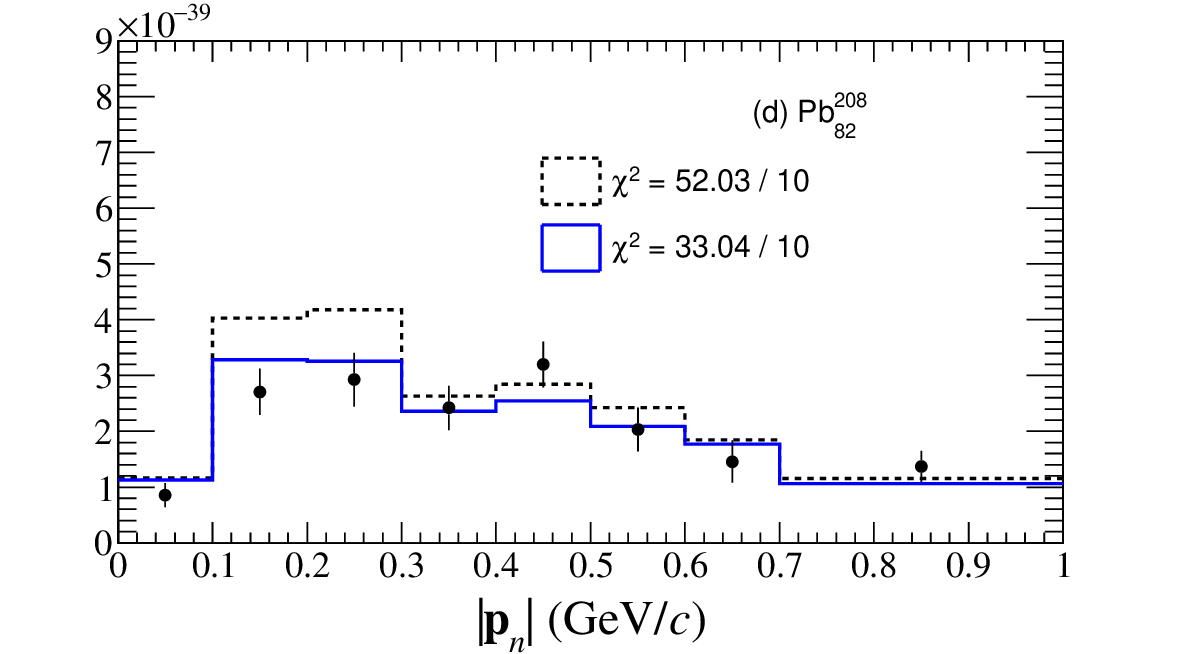}
    \end{subfigure}
    \caption{\label{fig:reconstructedNeutron_all_targets_cascde} (Color Online) Comparison of $|\mathbf{p}_{n}|$ with the \minerva \exclusivedata data..}
\end{figure*}

\par
Figures~\ref{fig:reconstructedNeutron_all_targets_cascde}(a)–(d) show the differential cross section measurements reported in~\cite{minerva-paper} as a function of $|\mathbf{p}_{n}|$ for carbon, water, iron, and lead. The peak at low values of $|\mathbf{p}_{n}|$ ($0$–$200$ MeV/$c$) reflects Fermi motion for CCQE events without \fsi. The region above $200$~MeV/c includes contributions from multi-nucleon knockout mechanism, pion absorption and CCQE with \fsi. In all the cases, we observe that increasing the strength of the nucleon-\fsi module improves the $\chi^{2}$/d.o.f. For carbon, improvements occur primarily in the first two bins, indicating fewer nucleon-nucleon re-interactions due to a light nucleus. Water shows a similar trend with improvements in the first three bins. As we go to heavier nuclei, the improvement in $\chi^{2}$/d.o.f becomes more pronounced. Lead exhibits the largest improvement, and the shapes produced by the two \nuwro configurations differ significantly. 
For lead, the peak height and tail thickness are comparable, indicating strong contributions from both Fermi motion and \fsi processes. 
\par Figures~\ref{fig:reconstructedNeutron_all_targets_cascde}(a)-(d) demonstrates that increasing the \fsi strength reduces the overall $|\mathbf{p}_{n}|$ cross section. This can be understood by analyzing how event-selection cuts affect the overall normalization. As the mean free path of propagating nucleons decreases, reinteractions become more frequent, shifting the momentum distribution of the leading proton toward lower values. Consequently, many events are removed to satisfy the muon and proton cut, see Eq.~(\ref{eqn:muon-cut}-\ref{eqn:proton-cut}), because they fall below the proton-momentum threshold of $|\mathbf{p}_{p}| = 0.5$ GeV/$c$, which leads to the lowering of the overall normalization. A measure to quantify the dilution of signal events caused by experimental cuts after increasing \fsi-strength is discussed in Ref.~\cite{PhysRevD.111.032009} for various experiments.

\par Even though the $\chi^2$/d.o.f improves after strengthening \nuwro's nucleon-\fsi effects the overall $\chi^2$/d.o.f is still rather poor. For carbon, oxygen, and iron, the primary contribution to the large values of $\chi^{2}$ originates from overestimation of the cross section in the first two bins. This is mainly due to our choice to model the initial nuclear state using the LFG model. This will be discussed in detail in Sect.~\ref{subsec:lfg_vs_sf}.\\

\begin{figure*}[htbp!]
    \begin{minipage}[b]{\columnwidth}
        \centering 
        \includegraphics[width=1.11\linewidth]{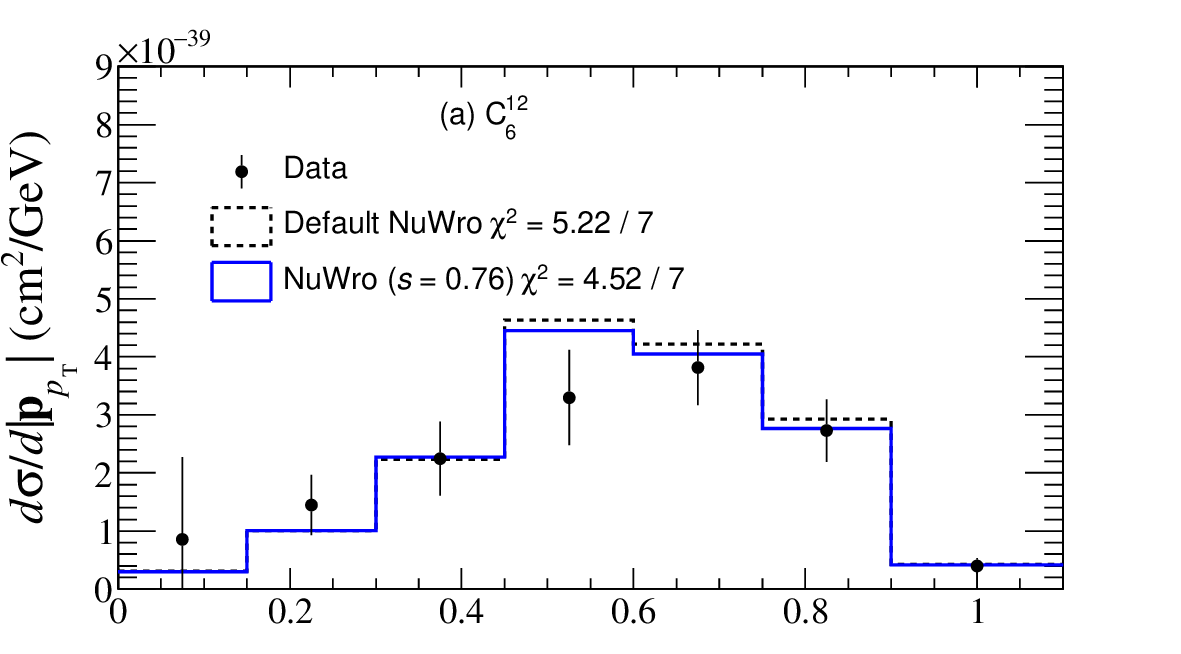}
    \end{minipage}
    \begin{minipage}[b]{\columnwidth}
        \centering 
        \includegraphics[width=1.11\linewidth]{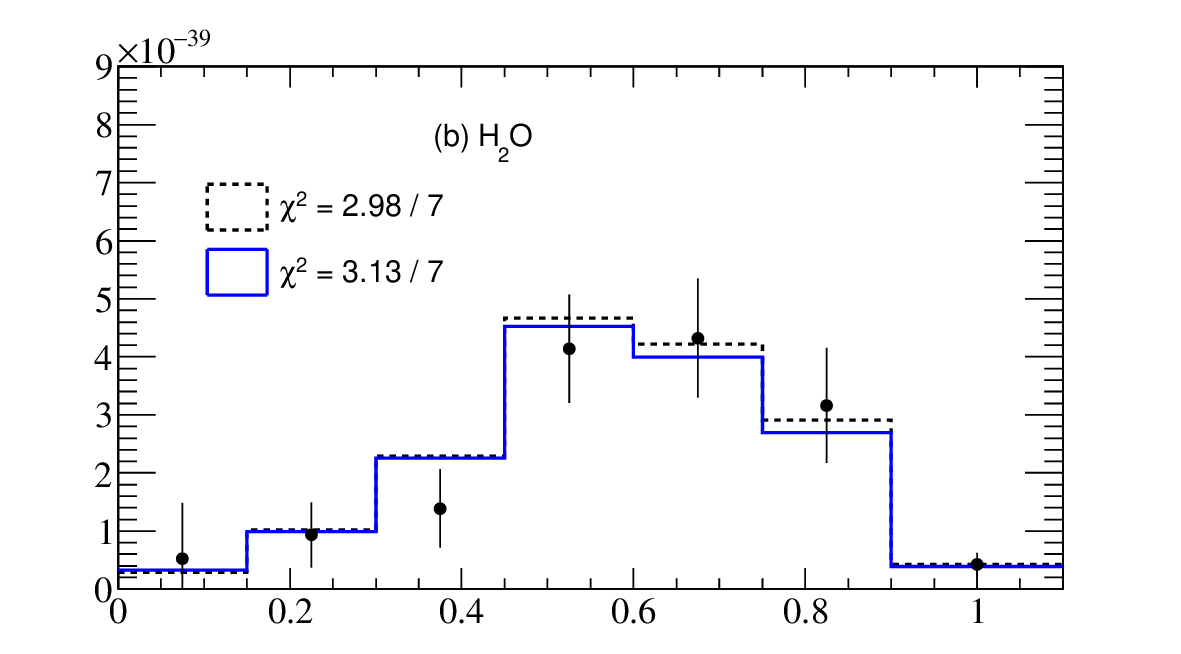}
    \end{minipage}
    \begin{minipage}[b]{\columnwidth}
        \centering 
        \includegraphics[width=1.11\linewidth]{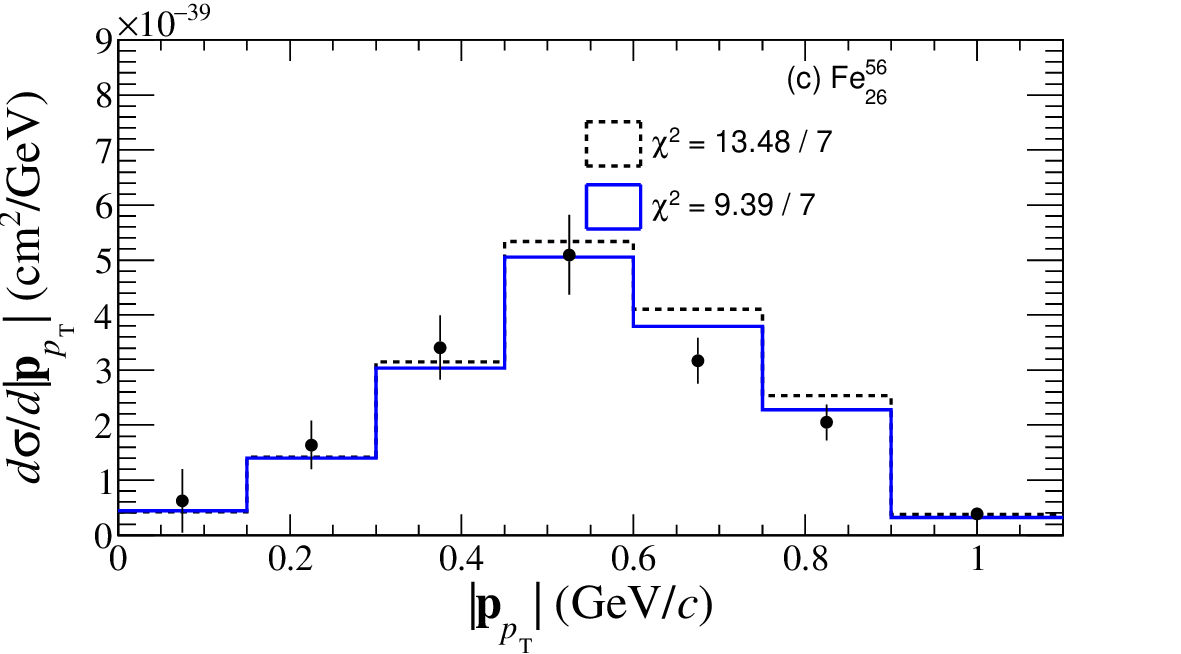}
    \end{minipage}
    \begin{minipage}[b]{\columnwidth}
        \centering 
        \includegraphics[width=1.11\linewidth]{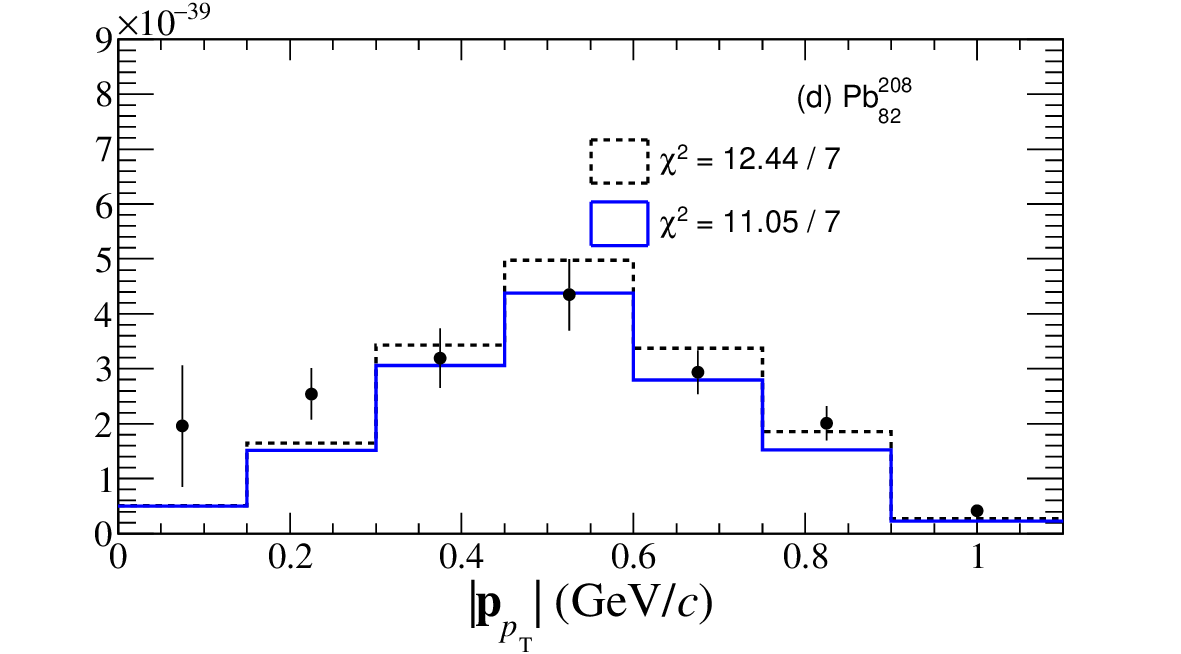}
    \end{minipage}
    \caption{\label{fig:transverseProtonMomentum_all_targets_cascde} (Color Online) Same as Fig.~\ref{fig:reconstructedNeutron_all_targets_cascde} but for $|\mathbf{p}_{p_{T}}|$}
\end{figure*}

\par
Figures~\ref{fig:transverseProtonMomentum_all_targets_cascde}(a)–(d) show the differential cross section measurements as a function of $|\mathbf{p}_{p_{T}}|$ for all nuclear targets. In Fig.~\ref{fig:transverseProtonMomentum_all_targets_cascde}(a), for carbon, we find that increasing the cascade strength changes the shape of the distribution only minimally. This occurs because most nucleons produced inside the carbon nucleus escape without significant re-interaction (see Sect.~\ref{sec:Introduction}) due to its smaller size. We observe a similar trend for water. Both carbon and water show good agreement with the data in terms of $\chi^{2}$/d.o.f., both before and after increasing the cascade strength. For iron, in Fig.~\ref{fig:transverseProtonMomentum_all_targets_cascde}(c), increasing the cascade strength reduces the number of events with $|\mathbf{p}_{p_{T}}| \gtrsim 0.5$ GeV, leading to improved agreement with the data. For lead, we observe that the cross section decreases in all bins except the first. The reason for this overall reduction is the same as in the case of $|\mathbf{p}_{n}|$. \\

A review of the remaining observables used in our analysis is moved to Appendix~\ref{sec:Appedices}. \\

\section{\label{sec:Discussion}DISCUSSION}
\subsection{\label{subsec:lfg_vs_sf}Sensitivity to Nuclear Modeling}

\par
In \nuwro, to describe nuclear effects in QE scattering on carbon, oxygen, argon, and iron targets, one can choose (among others) between the LFG and SF approaches. The SF approach is more realistic, but it is not available for lead.  In the LFG model, nuclear shell structure and nucleon correlations are absent. Nucleons are assumed to occupy all momenta up to the Fermi momentum $|\mathbf{p}_{F}(r)|$, which depends on the radial distance $r$ and the nuclear density, approximated in \nuwro using the charge distribution. In contrast, the SF approach incorporates shell structure and nucleon correlations~\cite{BENHAR1989267,BENHAR1994493}. It combines information from electron-scattering measurements of nuclear shells with theoretical calculations for nuclear matter at various densities.
\par Since the \minerva collaboration released correlation matrices for individual observables and four different nuclear targets together, the only consistent approach was to calculate \nuwro results with the same nuclear model, i.e., with LFG.\\

\begin{figure}[htbp!]
    \centering
    \begin{subfigure}[b]{\columnwidth}
    \includegraphics[width=\linewidth]{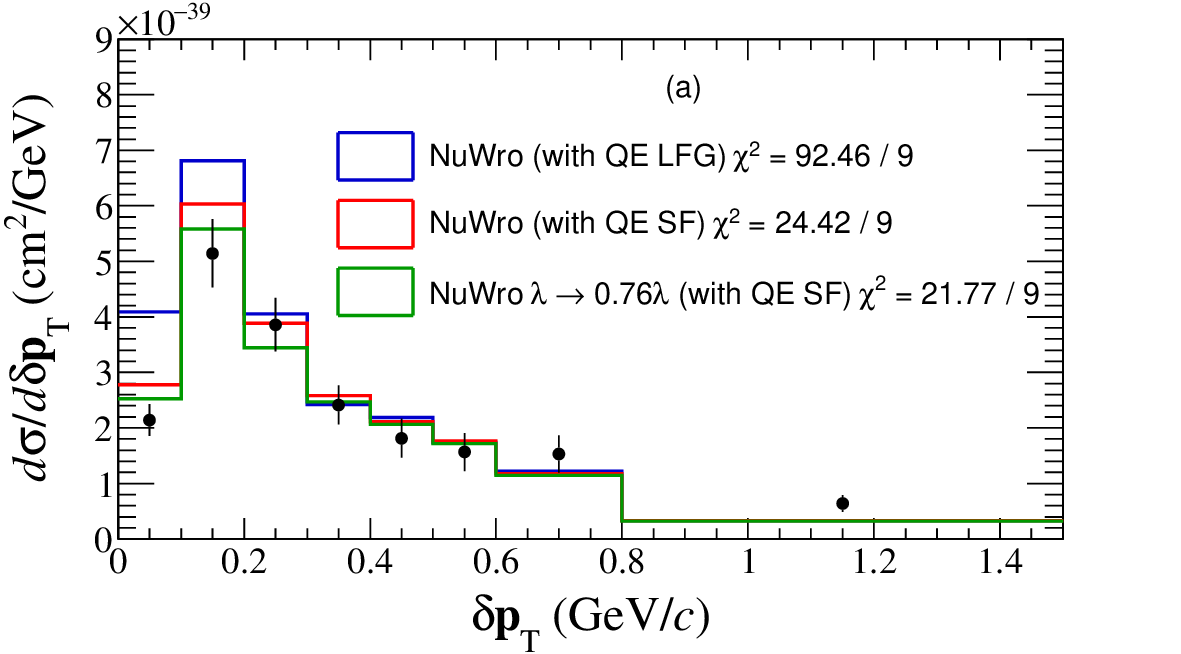}
    \end{subfigure}
    \hfill 
    \begin{subfigure}[b]{\columnwidth}
    \includegraphics[width=\linewidth]{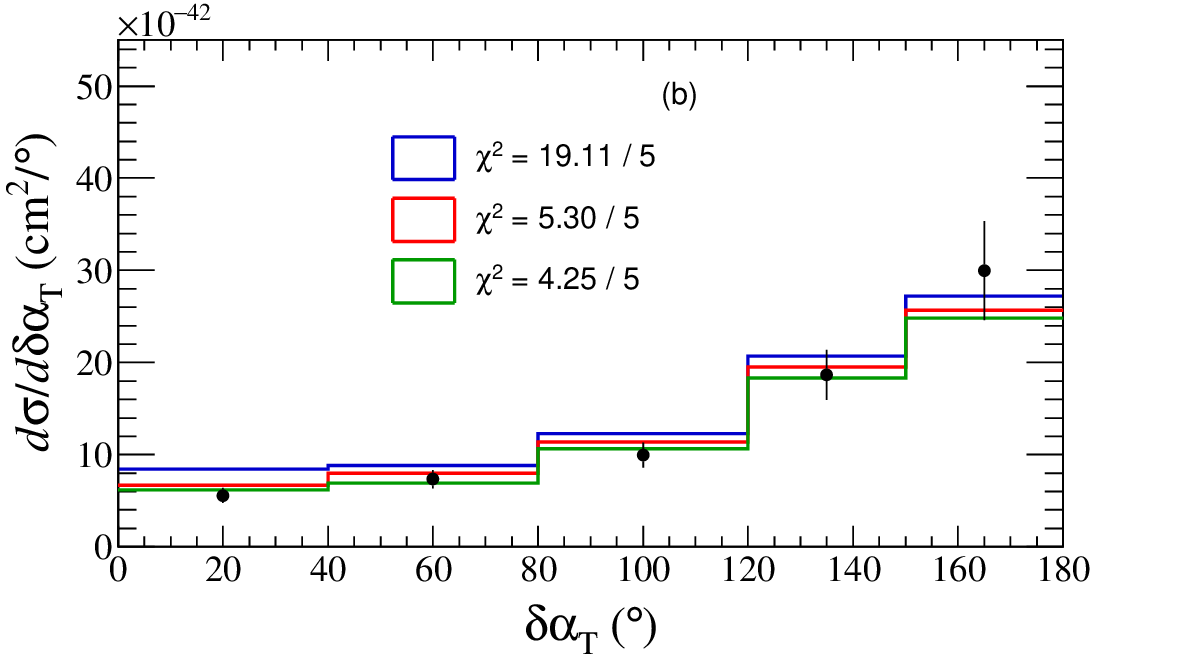}
    \end{subfigure}
    \hfill
    \begin{subfigure}[b]{\columnwidth}
    \includegraphics[width=\linewidth]{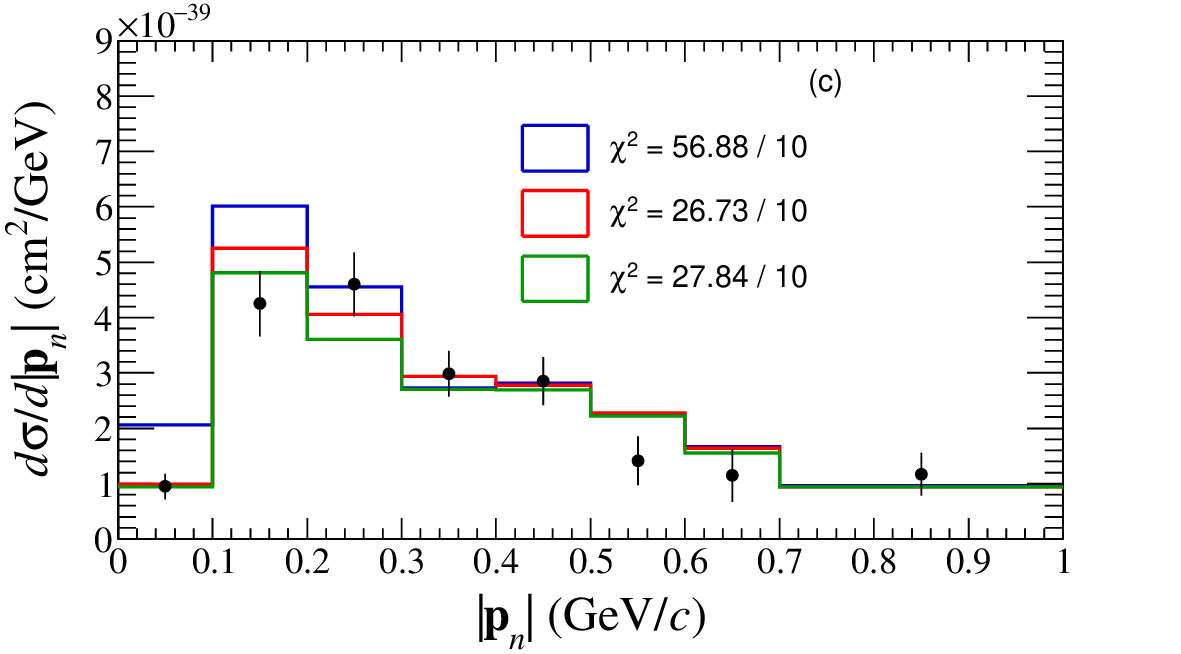}
    \end{subfigure}
    \caption{\label{fig:LFG-vs-SF} (Color Online) Same as Fig.~\ref{fig:reconstructedNeutron_all_targets_cascde} but for the iron target, comparing LFG and SF models for the initial nucleon state using $s = s_{0}$.}
\end{figure}

\par
It is essential to check if our findings are not biased by the use of an oversimplified nuclear model. Figures~\ref{fig:LFG-vs-SF}(a)–(c) show a comparison of \minerva\ measurements for iron with \nuwro predictions for \qe interactions simulated using the SF model at two cascade-strength values, along with predictions obtained using the LFG model at $s = 1$. In all cases, the LFG model overestimates the cross section. In the upper panel of Fig.~\ref{fig:LFG-vs-SF}, the LFG model significantly overestimates the first two bins—corresponding to the onset of the Fermi-motion peak—leading to poor $\chi^{2}$/d.o.f, whereas SF-based predictions show better agreement. Increasing the cascade strength further improves the agreement with data.\\

\par
In the middle panel of Fig.~\ref{fig:LFG-vs-SF}, although the shape of the $\delta\alpha_{T}$ distribution is similar for all approaches, SF produces a lower overall normalization, yielding improved $\chi^{2}$/d.o.f. Increasing the nucleon-\fsi strength provides a slight additional benefit.\\

\par The lower panel of Fig.~\ref{fig:LFG-vs-SF} also shows that the SF approach yields better $\chi^{2}$/d.o.f when compared to \nuwro (with QE LFG) predictions, demonstrating the sensitivity of \minerva-exclusive data to nuclear-state modeling. Only in the case of $|\mathbf{p}_{n}|$ we observe that \nuwro with $s = s_{0}$ and QE SF (histogram with green outline) the $\chi^{2}$/d.o.f is slightly worse compared to the $s = 1$ case (histogram with red outline). However, the difference between them is small, making it difficult to determine whether it reflects statistical fluctuations or indicates that the best-fit value of $s$ for this observable under the SF model is different. A tension between the second and third bins may indicate that the discrepancy is attributable to other components of the theoretical model (\mec and pion absorption). \\

\begin{table}[htbp!]
    \centering
    \caption{\label{tab:chi-square-table-SF} $\chi^{2}$/d.o.f.~values comparing \nuwro 
    predictions with the \minerva \exclusivedata data~\cite{minerva-paper} for iron, using the SF model to simulate \qe interactions.}
    \begin{ruledtabular}
    \begin{tabular}{ l | c | c }
      \textrm{Observable} & \nuwro($s=1$) & \nuwro($s=s_{0}$) \\
        \hline 
        1. $\delta\alpha_{T}$		     
        & $5.30 / 5$	& $4.25 / 5$ \\ 
        
        2. $\delta\mathbf{p}_{T}$        
        &	$24.42 / 9$	& $21.77 / 9$ \\
        
        3. $\delta\mathbf{p}_{T_{y}}$     
        & $20.11 / 9$ &	$9.07 / 9$ \\
        
         4. $|\mathbf{p}_{n}|$          
         &   $26.73 / 10$ & $27.84 / 10$ \\
         
         5. $\delta\phi_{T}$		      
         & $3.65 / 6$ &	$1.83 / 6$	\\ 
         
        6. $|\mathbf{p}_{p_{T}}|$       
        &   $5.52 / 7$ & $3.34 / 7$		\\
        
        7. $|\mathbf{p}_{p}|$ 
        & $17.51 / 5$	&    $14.57 / 5$ \\ 
        
        8. $\theta_{p}$		                 
        &  $3.98 / 10$ &	$4.72 / 10$ \\ 
    \end{tabular}
    \end{ruledtabular}
\end{table}

\par
Table~\ref{tab:chi-square-table-SF} lists the $\chi^{2}$/d.o.f for various observables for iron. Except for $|\mathbf{p}_{n}|$, all cases show reduced $\chi^{2}$ when the SF model is used. In instances where $\chi^{2} < \text{d.o.f.}$, increasing the cascade strength either improves agreement further or sometimes produces slightly worse $\chi^{2}$; the $\chi^2$/d.o.f remains below 1. Overall, we conclude that increasing the strength of the nucleon-\fsi module in \nuwro's cascade improves agreement with data, regardless of the nuclear model employed.

\subsection{\label{subsec:approx-scheme} \textit{Approximate} reweighting}

\par
To perform the exact reweighting described in Sect.~\ref{subsec:exact-scheme}, it is necessary to increase significantly the amount of information stored in each event. In this section, we explore a possibility to define an approximate reweighting method that yields a satisfactory approximation of the exact process, while requiring less storage. The goal is to develop a tool that could be easily used in future MC \fsi studies.\\

\par 
The main idea in this approach is to use approximate values of (non-)interaction probability. In a typical event simulated by \nuwro's \fsi module, the total number of steps ($N = N_S + N_f$), for an intermediate size nucleus, is about $\sim 1000$, while the probability of interaction (p) is around $\sim 0.01$. We can estimate the average interaction probability ($\hat{\text{p}}$) using pre-generated samples from \nuwro as:
\begin{equation}
    \label{eqn:mle-binomial}
    \hat{\text{p}} = \left( \sum_{i=1}^{M} N_{s_i} \right) \bigg/ \left( \sum_{i=1}^{M} N_i \right)
\end{equation}
where $M$ is the number of events, and $N_i = N_{s_i} + N_{f_i}$ is the overall number of steps for the $i^{\text{th}}$ event. Table~\ref{tab:probability} shows the values of $\hat{\text{p}}$ obtained for different nuclear targets using the corresponding NuMI ME flux.
\begin{table}[htbp!]
\centering
\caption{\label{tab:probability} Average interaction probability $\hat{\text{p}}$ for different nuclear targets using the corresponding NuMI ME flux~\cite{minerva-paper} using $10^{5}$ events. }
\begin{tabular}{ | l | c | }
\hline 
Nuclear target & $\hat{\text{p}}$ \\
\hline 
1. Carbon & $8.04\times 10^{-3}$\\
2. Water  & $8.05\times 10^{-3}$\\
3. Iron  & $1.38\times 10^{-2}$\\
4. Lead & $2.20\times 10^{-2}$\\
\hline
\end{tabular}
\end{table}

The estimator $\hat{\text{p}}$ represents the probability of interaction at each step, within the nucleus. In the approximate approach, the problem is to express the scaling of the average interaction probability, which differs across nuclear targets, in terms of the \textit{universal} rescaling of the mean-free path, which is the same for all targets. With this information, we scale the estimator $\hat{\text{p}}$, as:
\begin{equation}
    \label{eqn:scaling-interaction-probability-approximate}
    \hat{\text{p}} \to \hat{\text{p}}' = 1 - (1-\hat{\text{p}})^{1/r}
\end{equation}
where $r$ will be the the scaling parameter in the \textit{approximate} scheme, analogous to $s$ parameter in the case of Algorithm~\ref{alg:exact_reweighting_scheme}.

\par
\begin{figure}[htbp!]
    \centering
    \includegraphics[width=\linewidth]{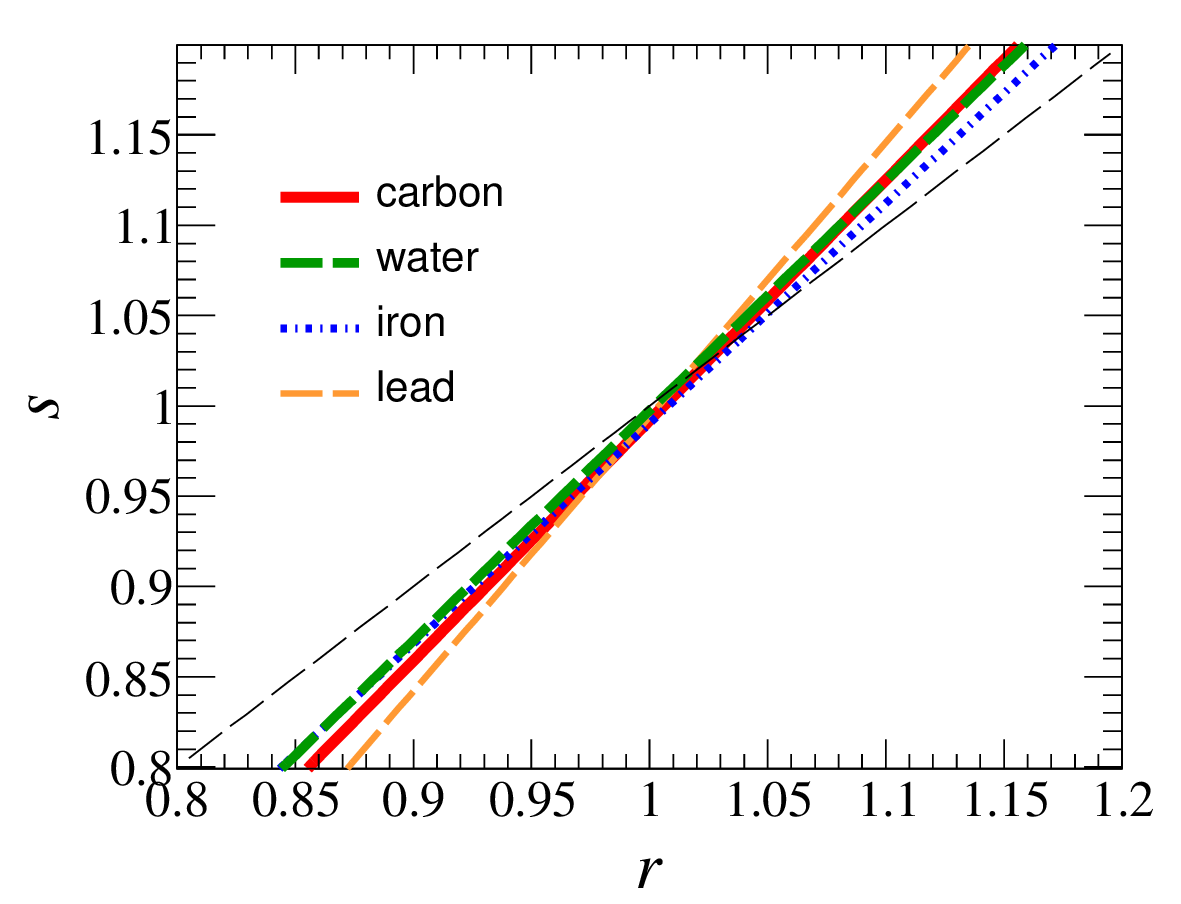}
    \caption{A relationship between the scaling parameter $r$ for interaction probability $\hat{\text{p}}$ and the scaling parameter $s$ for mean free path.}
    \label{fig:dictParameter}
\end{figure}

 We attempted to relate $r$ to the \textit{universal} rescaling parameter $s$. To do this, we fixed a nuclear target and produced a \nuwro output file with the mean free path scaled by a value $s=s'$. We then reweighted events from a default \nuwro  output file ($s = 1$) using the reweight parameter $r$ until the momentum distribution of the most energetic proton matched the distribution generated by \nuwro with $s = s'$ within a 1\% tolerance; we denote this value of $r$ by $r'$. We then say that scaling the average interaction probability $\hat{\text{p}}$ by $r'$ produces similar effects to scaling the mean free path by $s$. We then repeat this procedure for a sufficient number of $s'$ values and obtain a linear fit of $s$ versus $r$. Figure~\ref{fig:dictParameter} shows the resulting relation between $r$ and $s$ obtained with this method for different nuclear targets.\\

\begin{algorithm}[htbp!]
    \caption{\label{alg:approximate_reweighting_scheme} \textit{Approximate}-reweighting scheme}

    \KwInput{\nuwro output \texttt{.root} file with $M$ events.}
    
    Fix the value of the \textit{reweight parameter} $r$.
    
    Scale the average probability of interaction: $\hat{\text{p}}$ using Eq.~(\ref{eqn:scaling-interaction-probability-approximate}).

    \For{event $i = 1$ \KwTo $M$}{

        Assign new weight $w_{i}(r) =  {\displaystyle \mathcal{L} (N_{s_i}, N_{f_i} \,;\, \hat{\text{p}}')\over \displaystyle \mathcal{L} (N_{s_i}, N_{f_i} \,;\, \hat{\text{p}})} $
    
    }
    \KwResult{New \texttt{.root} file with weighted events.}
\end{algorithm}

\par 
In this scheme, we define the likelihood of the $i^{\text{th}}$ event as:
\begin{equation}
    \label{eqn:likelihood-approx}
    \mathcal{L} (N_{s_i}, N_{f_i} \,;\, \hat{\text{p}}) \approx \displaystyle (1-\hat{\text{p}})^{N_{f_i}}\,\displaystyle\hat{\text{p}}^{N_{s_i}}
\end{equation}
After scaling the interaction probability, $\hat{\text{p}}$, the new likelihood of the event is $\mathcal{L} (N_{s_i}, N_{f_i} \,;\, \hat{\text{p}}')$ where $\hat{\text{p}}'$ is the scaled interaction probability. The reweighting factor $w_{i}(r)$ is given as:
\begin{equation}
    w_{i}(r) =  {\displaystyle \mathcal{L} (N_{s_i}, N_{f_i} \,;\, \hat{\text{p}}')\over \displaystyle \mathcal{L} (N_{s_i}, N_{f_i} \,;\, \hat{\text{p}})} .
\end{equation}

\par We quantify the performance of the \textit{approximate}-reweighting scheme using the quantitative measure $\epsilon(r)$ (same as $\epsilon(s)$ in Eq.~(\ref{eqn:scheme-performance})). For this, we produced a sample of $10^{5}$ events using the \nuwro MC generator (with $s=1$) and reweighted events for a range of $r$. 
\begin{figure}[htbp!]
    \centering
    \includegraphics[width=\linewidth]{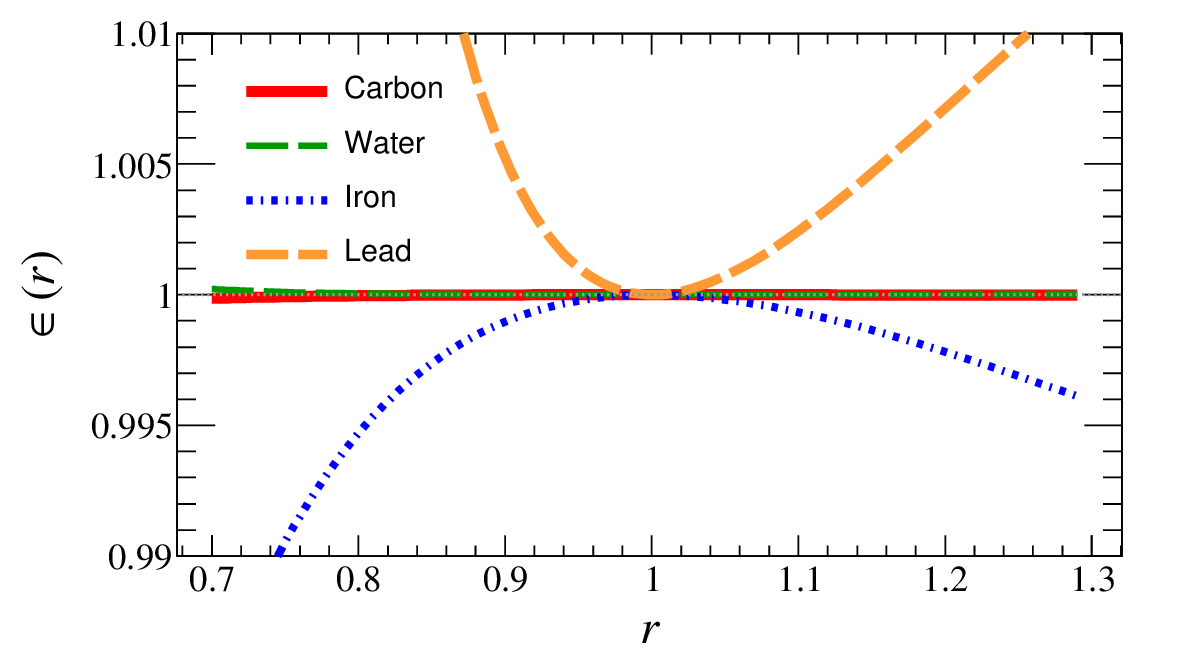}
    \caption{\label{fig:approx-scheme-performance} (Color online) Effect on the overall normalization using the measure $\epsilon(r)$ for the method given in Algorithm~\ref{alg:approximate_reweighting_scheme} for various nuclear targets.}
\end{figure}
In Fig.~\ref{fig:approx-scheme-performance}, we quantify the performance of the \textit{approximate}-scheme in terms of preservation of the overall normalization. The normalization is maintained with perfect precision for carbon and oxygen. For iron, the normalization remains stable within 1\% in the interval $[0.75,\,1.3]$. For the lead, the situation is worse, as normalization is preserved at most up to 1\% within the interval $[0.87, 1.26]$. 

\par We then examined whether the differential cross sections, expressed in terms of lepton kinematic variables, remain stable when reweighting within statistical fluctuations. For the sample with $10^{5}$ events, this condition is typically satisfied over the range of  $[0.90, 1.11]$.\\

\begin{figure}[htbp!]
    \centering
    \includegraphics[width=\linewidth]{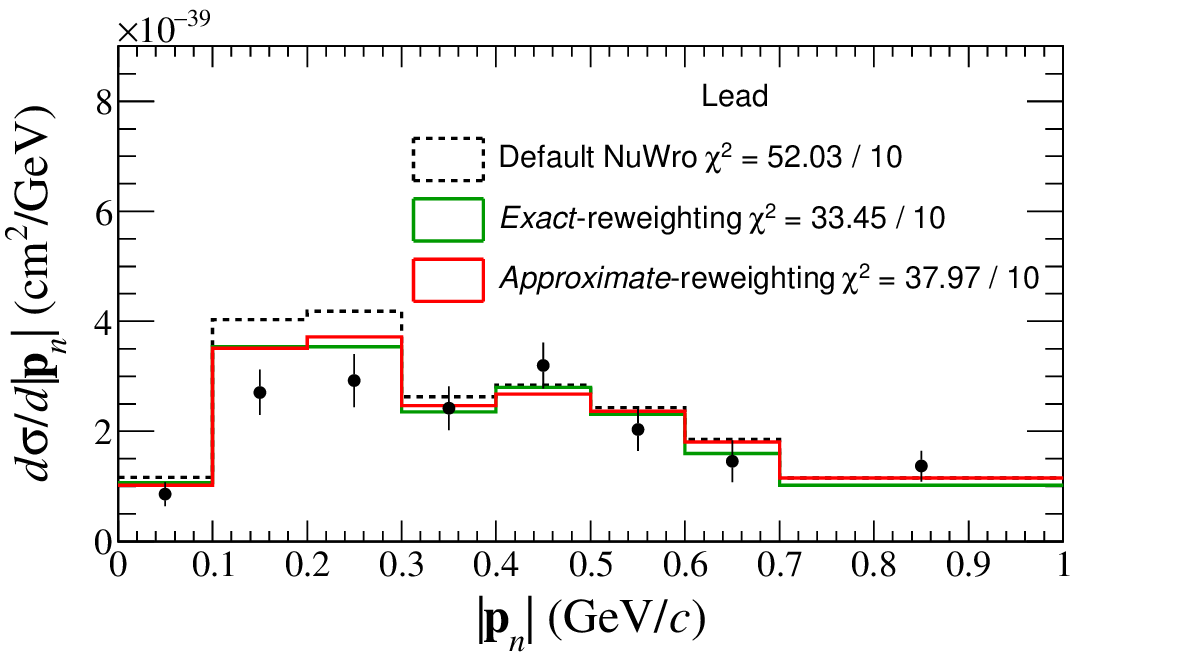}
    \caption{\label{fig:exact-vs-approx} A comparison between the \textit{exact}- and \textit{approximate}-reweighting schemes for $r=0.90$. Data points are from ~\cite{minerva-paper}.}
\end{figure}

\par We qualitatively compare the \textit{exact} and \textit{approximate} reweighting schemes in Fig.~\ref{fig:exact-vs-approx}. We choose $r=0.9$, which lies within the region of applicability of both the \textit{approximate} and \textit{exact} schemes. The shapes produced by the two schemes are very similar, demonstrating the robustness of the \textit{approximate} scheme, which uses only a fraction of the information compared with the \textit{exact} scheme. The $\chi^{2}$/d.o.f obtained with the \textit{exact} and \textit{approximate} reweighting scheme are similar, which proves that the \textit{approximate} scheme may be a useful tool that helps to get a general orientation on how modification of the strength of \fsi modifies \nuwro predictions. \\

\par We also sought to relax certain assumptions while retaining the information required to reweight an \fsi event, thereby improving the performance of the \textit{approximate} scheme. In this scheme, it is assumed that the nucleon interaction probability (success rate) is constant throughout the nuclear volume. However, this is valid only if the overall nuclear density is constant. In reality, the success rate varies with the positions of re-interaction points within the nucleus. There is a relatively high success rate when the interaction point is deep within the nucleus, and much lower at the outskirts of the nucleus, where the density is minimal. Based on the nuclear density profile of a given nucleus, we tried to divide the total volume of the nucleus into three different zones. We then counted successes and failures in each zone using Eq.~(\ref{eqn:mle-binomial}) and computed three separate average interaction probabilities. In this case, we defined the likelihood of an event as:
\begin{equation}
    \label{eqn:likelihood-approx-3zones}
    \mathcal{L} (N_{s_i}, N_{f_i} \,;\, \hat{\text{p}}_{\text{I}}, \hat{\text{p}}_{\text{II}}, \hat{\text{p}}_{\text{III}}  ) \approx \displaystyle\prod_{\text{zone}}^{\{\text{I, II, III}\}} \displaystyle(1-\hat{\text{p}}_{\text{zone}})^{N_{f_i}^{\text{zone}}}\,\displaystyle\hat{\text{p}}_{\text{zone}}^{N_{s_i}^{\text{zone}}}
\end{equation}
where $N_{s_i}, N_{f_i}$ are the total number of interaction steps and non-interaction steps in three zones. We then scaled the interaction probability in each zone using Eq.~(\ref{eqn:scaling-interaction-probability-approximate}) and assigned the additional weight to the event, defined as the ratio of the likelihood $W_{i}(r) = {\displaystyle \mathcal{L}' \over \displaystyle \mathcal{L}}$. We concluded that the performance gain was smaller than expected.\\

\section{\label{sec:Conclusion}CONCLUSIONS}
\par In this article, we fine-tune the strength of \nuwro's nucleon-\fsi part of the cascade module using recent \exclusivedata data~\cite {minerva-paper} from the \minerva experiment, performed simultaneously on carbon, oxygen, iron, and lead. For this, we developed an \textit{exact}-reweighting framework to reweigh events with nucleon FSI in \nuwro without re-running the whole simulation. The procedure is based on scaling the probability of (non-)interactions of the propagating nucleon(s) in a well-defined manner and quantifying the occurrence of each event by computing its likelihood. \\ 

\par Using \minerva's \exclusivedata measurements on carbon, water, iron, and lead, we performed a global fit of several exclusive observables which are sensitive to nuclear effects like \fsi, including TKI variables, reconstructed neutron momentum, and proton transverse momentum. The joint-likelihood combinations of all observables used in our analysis indicated a consistently preferred increase in \fsi strength by:

\begin{equation}
    s_{0} = 0.76^{-0.04}_{+0.06}
\end{equation}

The new uncertainty band around $s_0$ is around $\sim 10\%$. The tuned \fsi parameter $s$ and the reweighting implementation presented here will be incorporated into the next \nuwro release, providing a more realistic estimate of \fsi uncertainties and improving generator predictions for current and future oscillation experiments.\\

\section{\label{sec:Acknowledgement}ACKNOWLEDGMENT}
\par 
We would like to thank Clarence Wret and Kamil Skwarczynski for
providing us with details of the implementation of \fsi-reweighting in \neut. This work was partly (A.M.A.,
K.M.G., B.E.K., and J.T.S.) or fully (R.D.B., J.L.B., and
H.P.) supported by the National Science Centre under
grant UMO-2021/41/B/ST2/02778. K.M.G. is also partly supported by the Excellence Initiative—Research University, 2020–2026 at the University of Wroclaw.
\\

\appendix
\section{\label{sec:Appedices} Impact of reweighting \fsi on differential cross sections}

\begin{figure*}[hbp!]
    \begin{minipage}[b]{\columnwidth}
        \centering 
        \includegraphics[width=1.05\linewidth]{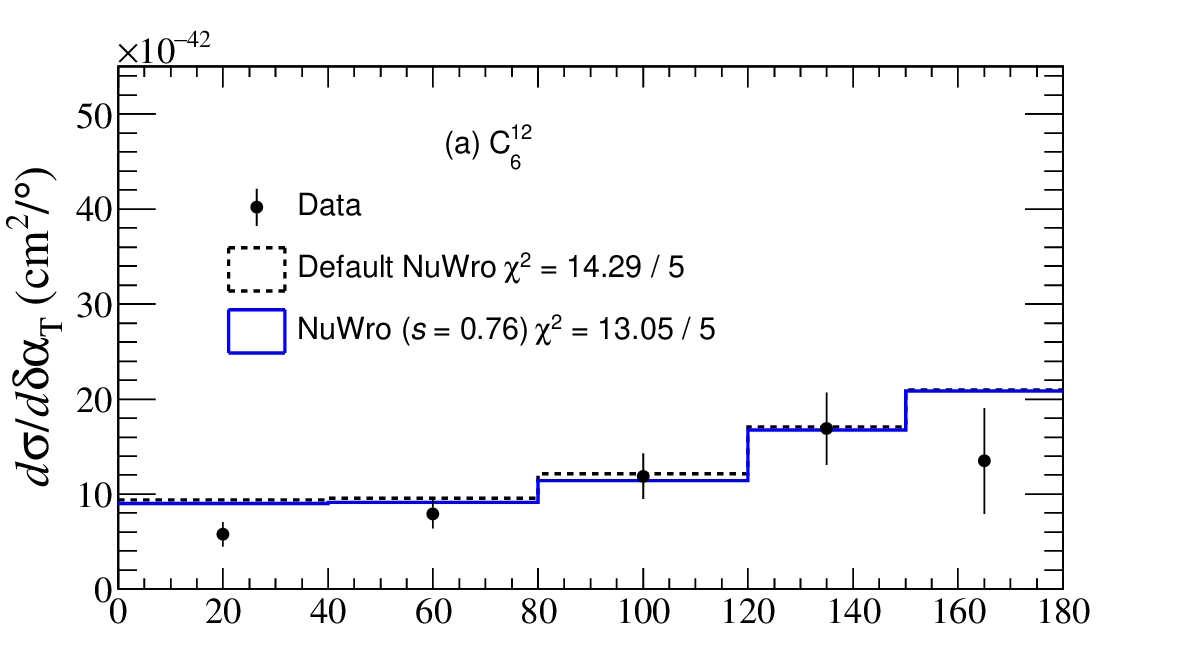}
    \end{minipage}
    \begin{minipage}[b]{\columnwidth}
        \centering 
        \includegraphics[width=1.05\linewidth]{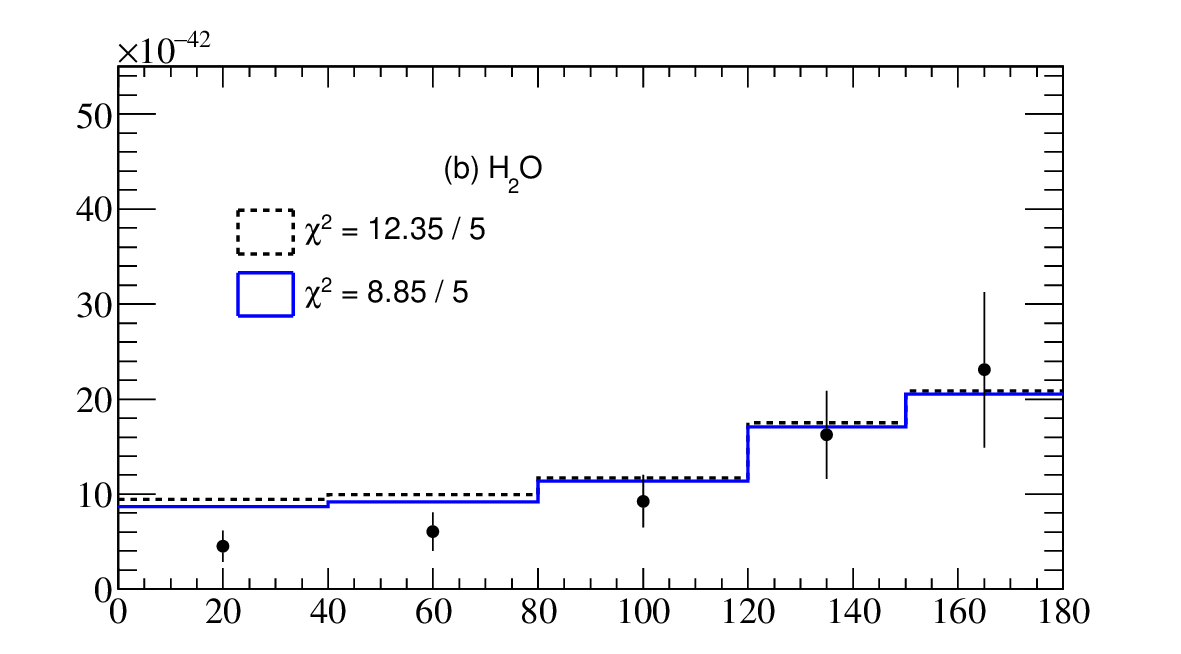}
    \end{minipage}
    \begin{minipage}[b]{\columnwidth}
        \centering 
        \includegraphics[width=1.05\linewidth]{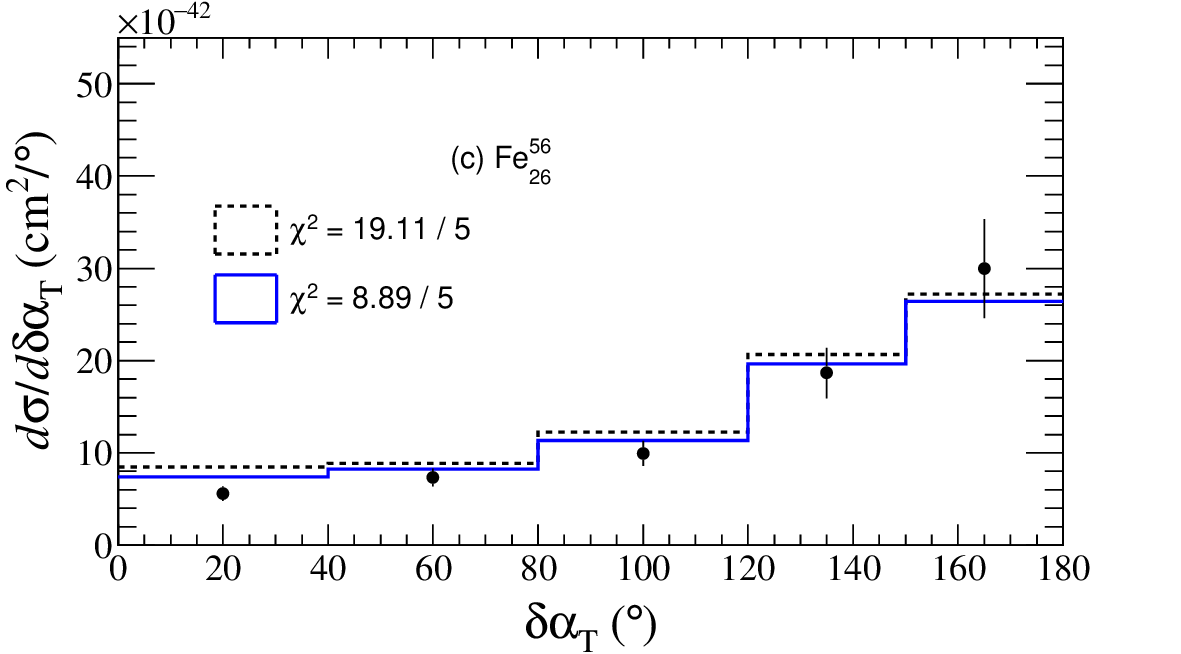}
    \end{minipage}
    \begin{minipage}[b]{\columnwidth}
        \centering 
        \includegraphics[width=1.05\linewidth]{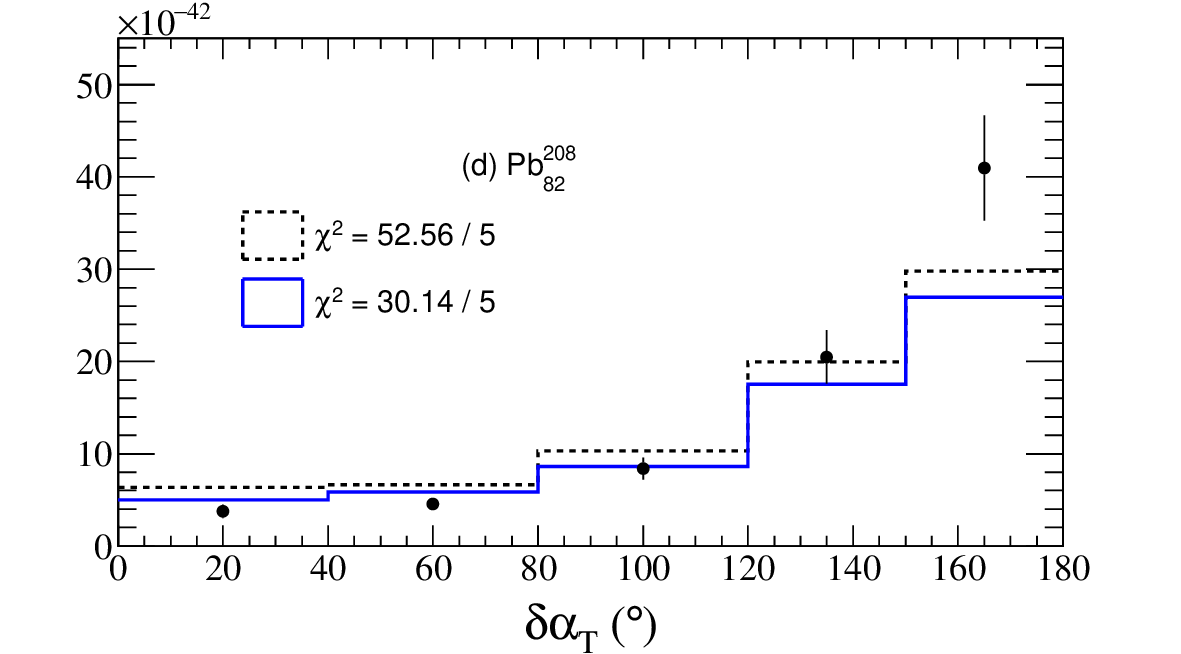}
    \end{minipage}
    \caption{\label{fig:dalphat_all_targets_cascade} (Color Online) Same as Fig.~\ref{fig:reconstructedNeutron_all_targets_cascde} but for $\delta\alpha_{T}$.}
\end{figure*}

\begin{figure*}[htbp]
    \begin{subfigure}[b]{\columnwidth}
        \centering 
        \includegraphics[width=1.05\linewidth]{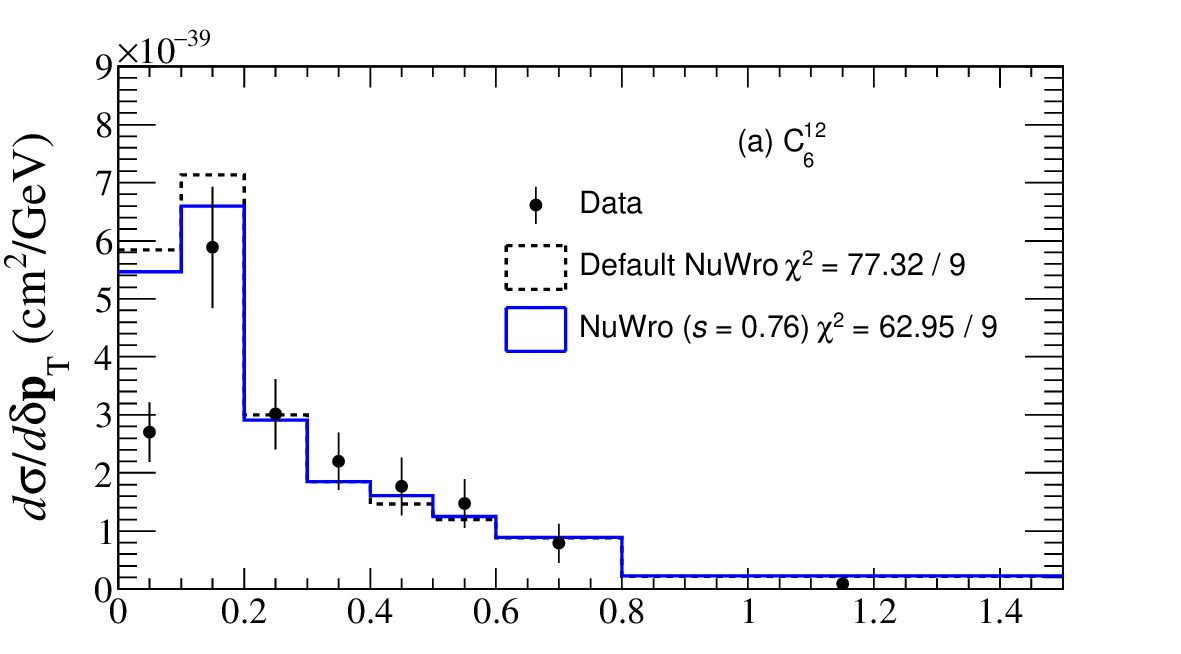}
    \end{subfigure}
    \begin{subfigure}[b]{\columnwidth}
        \centering 
        \includegraphics[width=1.05\linewidth]{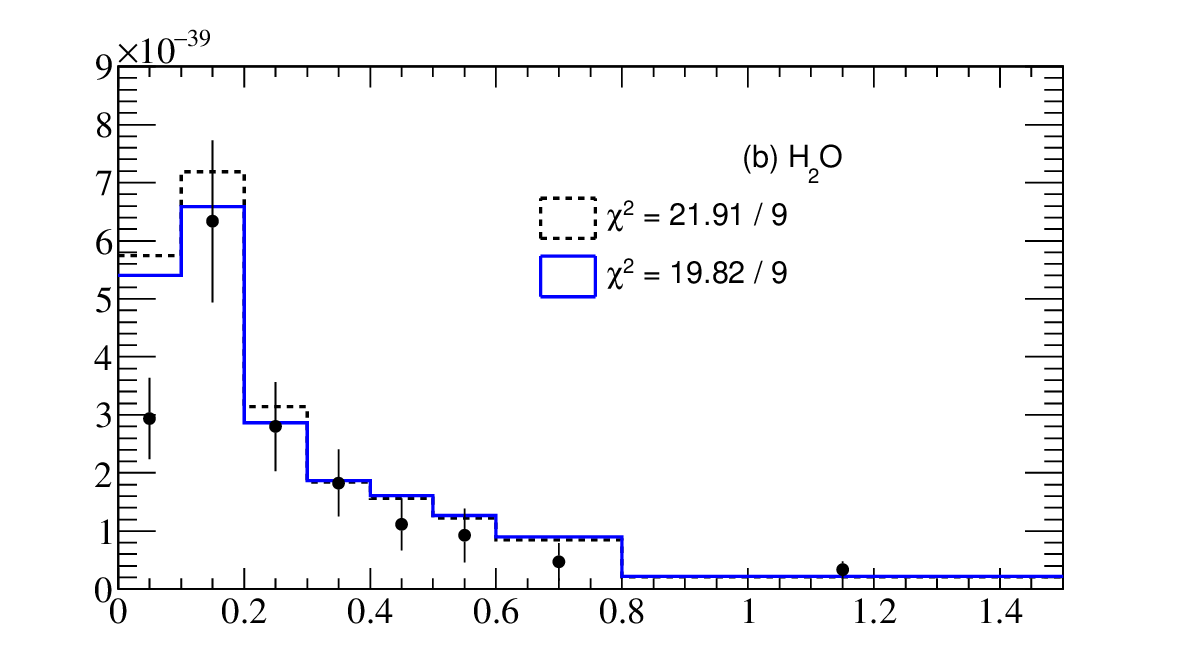}
    \end{subfigure}
    \begin{subfigure}[b]{\columnwidth}
        \centering 
        \includegraphics[width=1.05\linewidth]{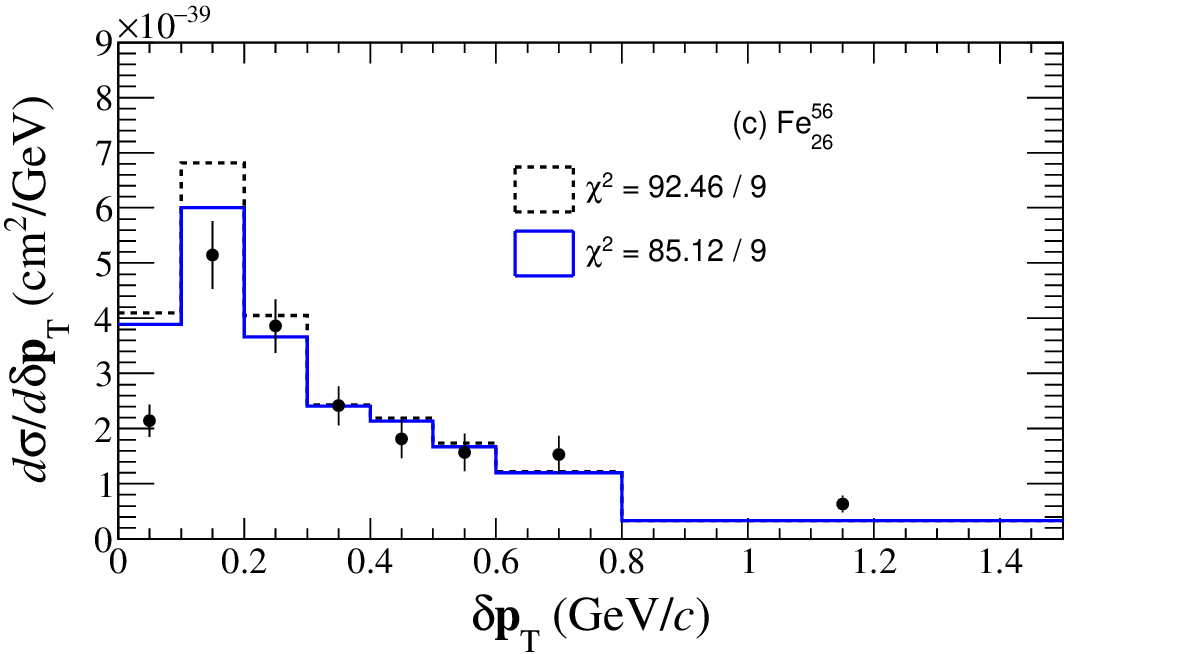}
    \end{subfigure}
    \begin{subfigure}[b]{\columnwidth}
        \centering 
        \includegraphics[width=1.05\linewidth]{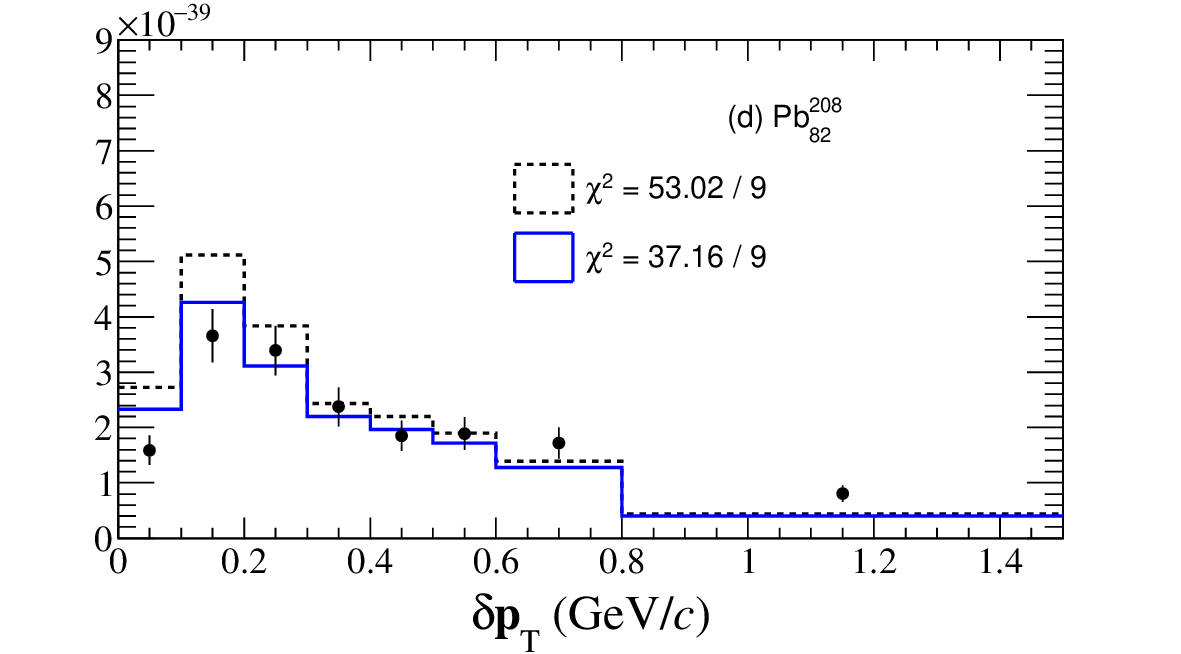}
    \end{subfigure}
    \caption{\label{fig:dpt_all_targets_cascade} (Color Online) Same as Fig.~\ref{fig:reconstructedNeutron_all_targets_cascde} but for $|\delta\mathbf{p}_{T}|$.}
\end{figure*}

\begin{figure*}[htbp!]
    \begin{minipage}[b]{\columnwidth}
        \centering 
        \includegraphics[width=1.05\linewidth]{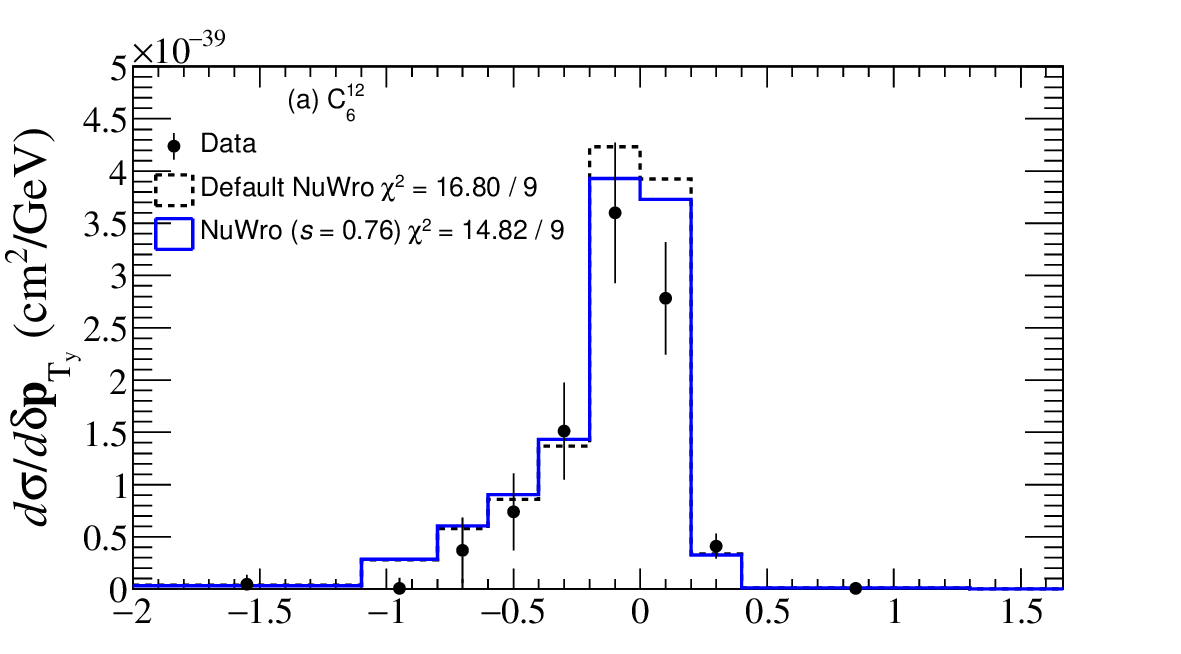}
    \end{minipage}
    \begin{minipage}[b]{\columnwidth}
        \centering 
        \includegraphics[width=1.05\linewidth]{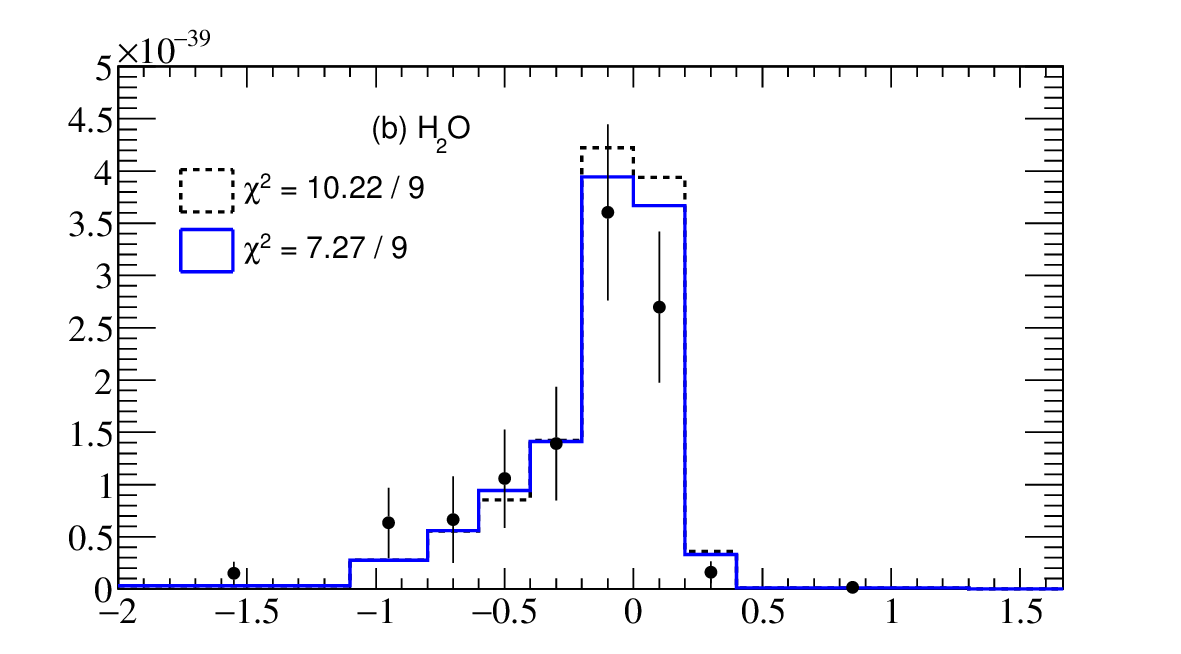}
    \end{minipage}
    \begin{minipage}[b]{\columnwidth}
        \centering 
        \includegraphics[width=1.05\linewidth]{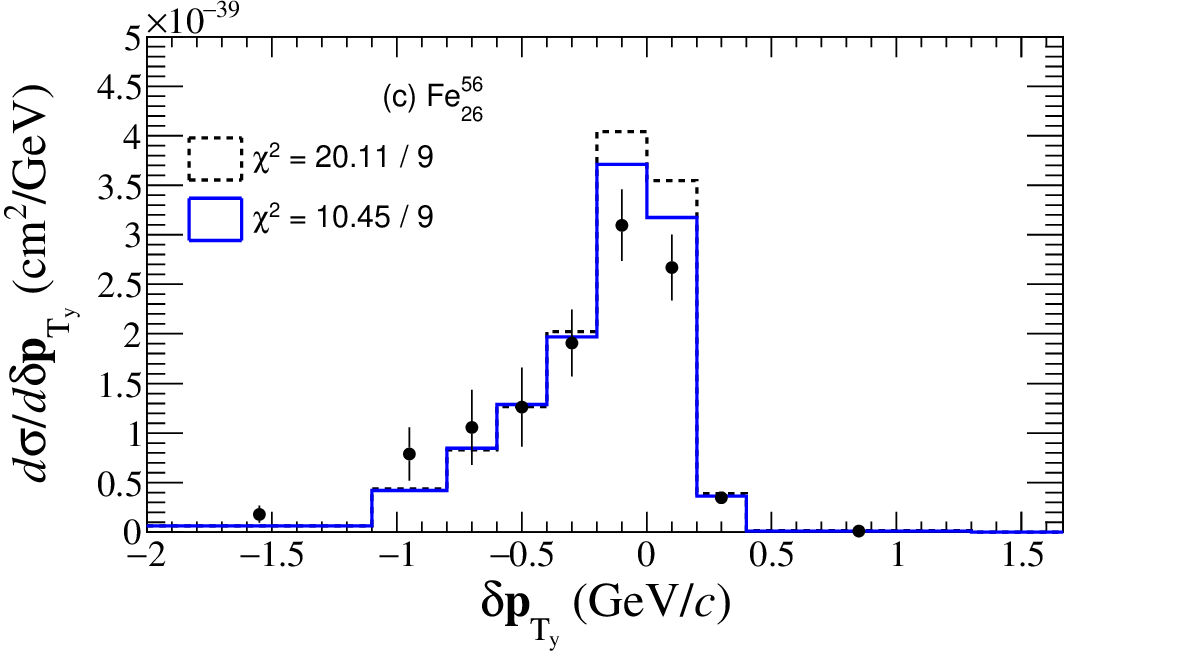}
    \end{minipage}
    \begin{minipage}[b]{\columnwidth}
        \centering 
        \includegraphics[width=1.05\linewidth]{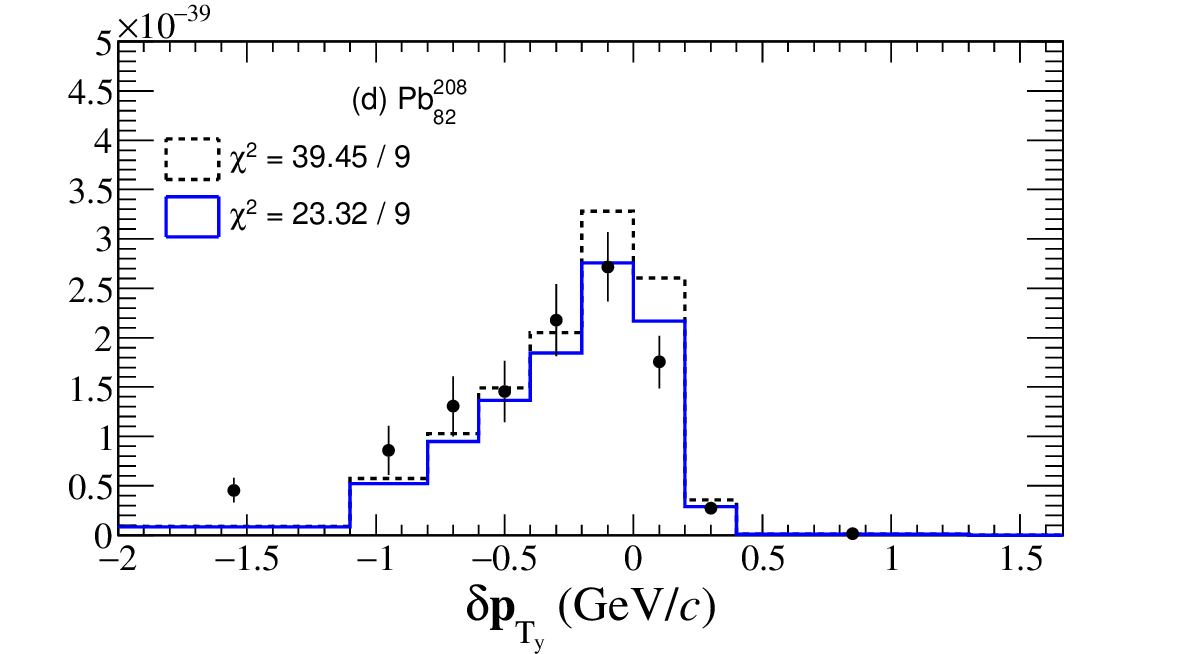}
    \end{minipage}
    \caption{\label{fig:dpty_all_targets_cascde} (Color Online) Same as Fig.~\ref{fig:reconstructedNeutron_all_targets_cascde} but for $\delta\mathbf{p}_{T_{y}}$.}
\end{figure*}

\begin{figure*}[htbp!]
    \begin{minipage}[b]{\columnwidth}
        \centering 
        \includegraphics[width=1.05\linewidth]{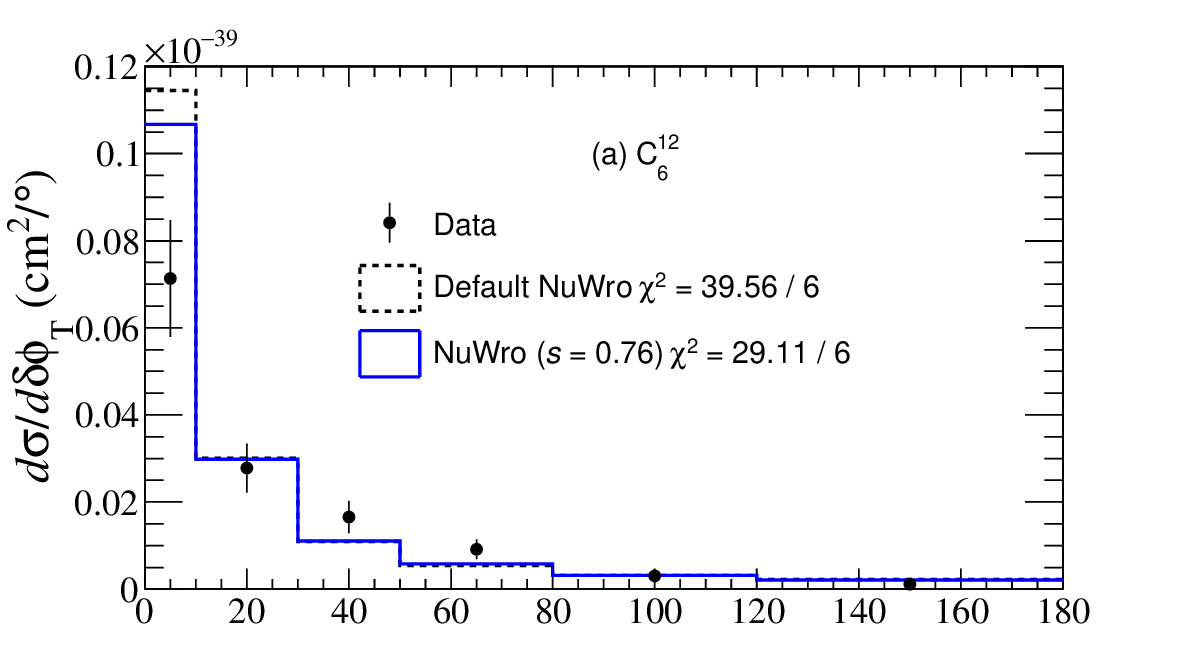}
    \end{minipage}
    \begin{minipage}[b]{\columnwidth}
        \centering 
        \includegraphics[width=1.05\linewidth]{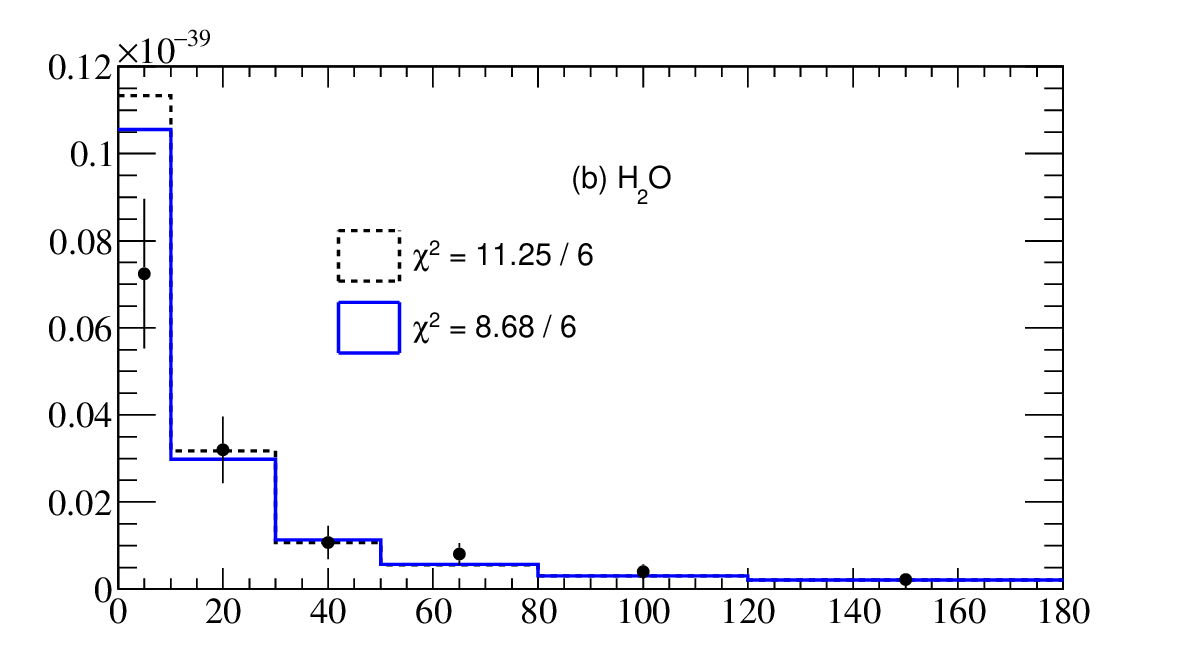}
    \end{minipage}
    \begin{minipage}[b]{\columnwidth}
        \centering 
        \includegraphics[width=1.05\linewidth]{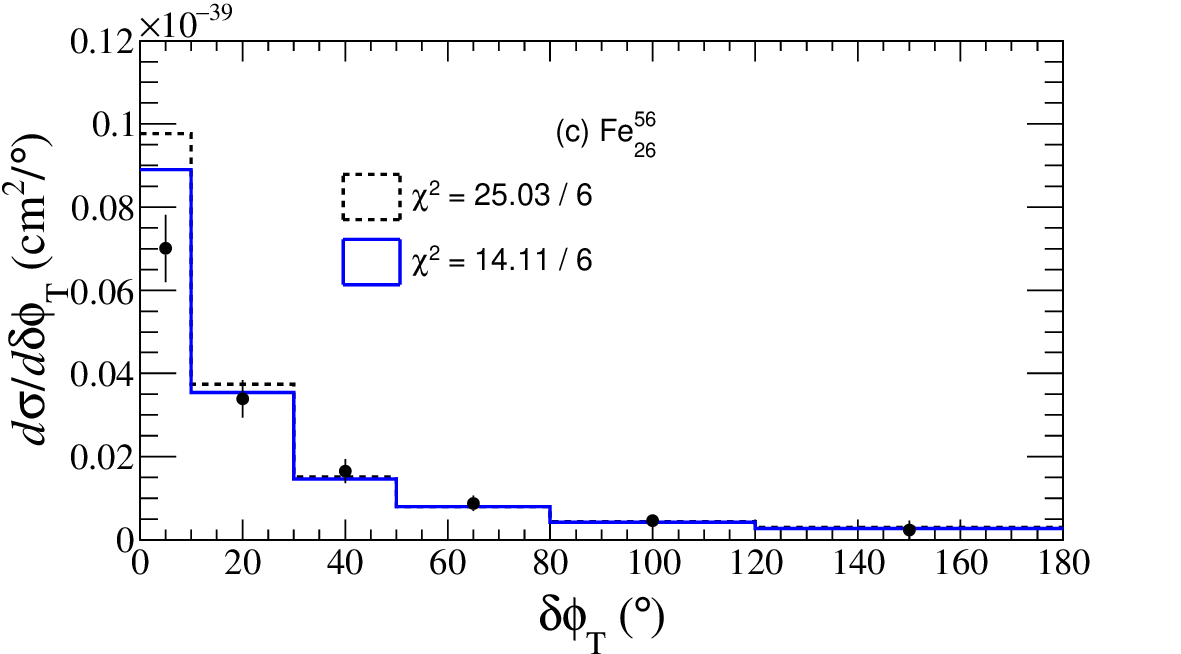}
    \end{minipage}
    \begin{minipage}[b]{\columnwidth}
        \centering 
        \includegraphics[width=1.05\linewidth]{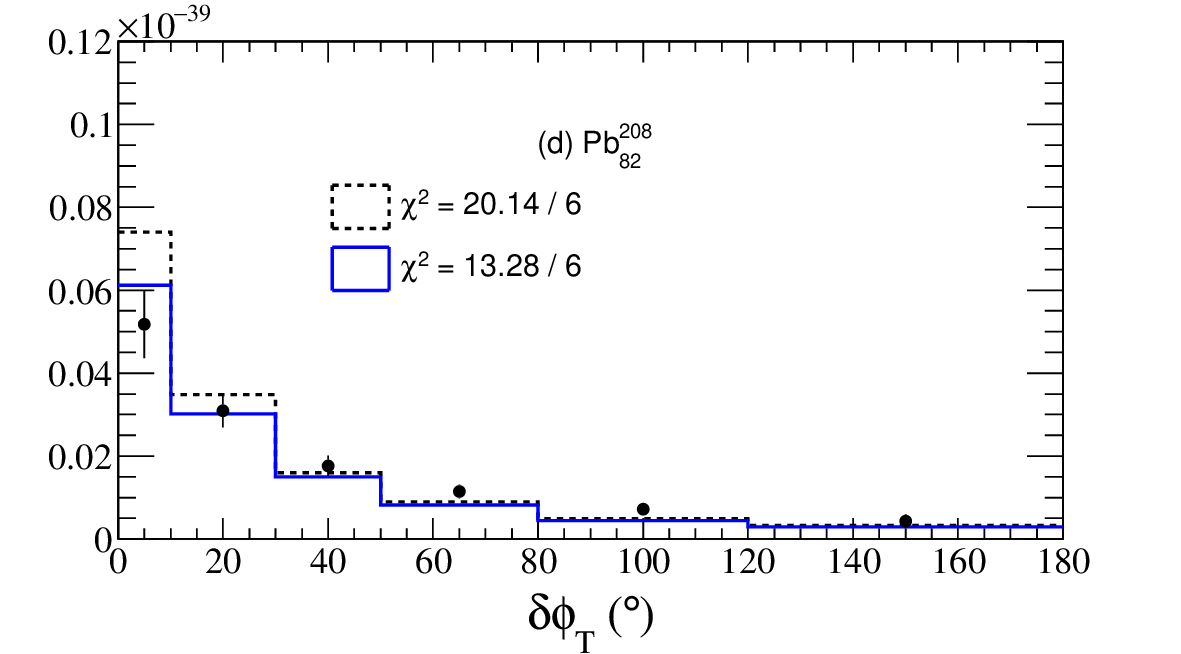}
    \end{minipage}
    \caption{\label{fig:dphit_all_targets_cascade} (Color Online) Same as Fig.~\ref{fig:reconstructedNeutron_all_targets_cascde} but for $\delta\phi_{T}$ }
\end{figure*}

\begin{figure*}[htbp!]
    \begin{minipage}[b]{\columnwidth}
        \centering 
        \includegraphics[width=1.05\linewidth]{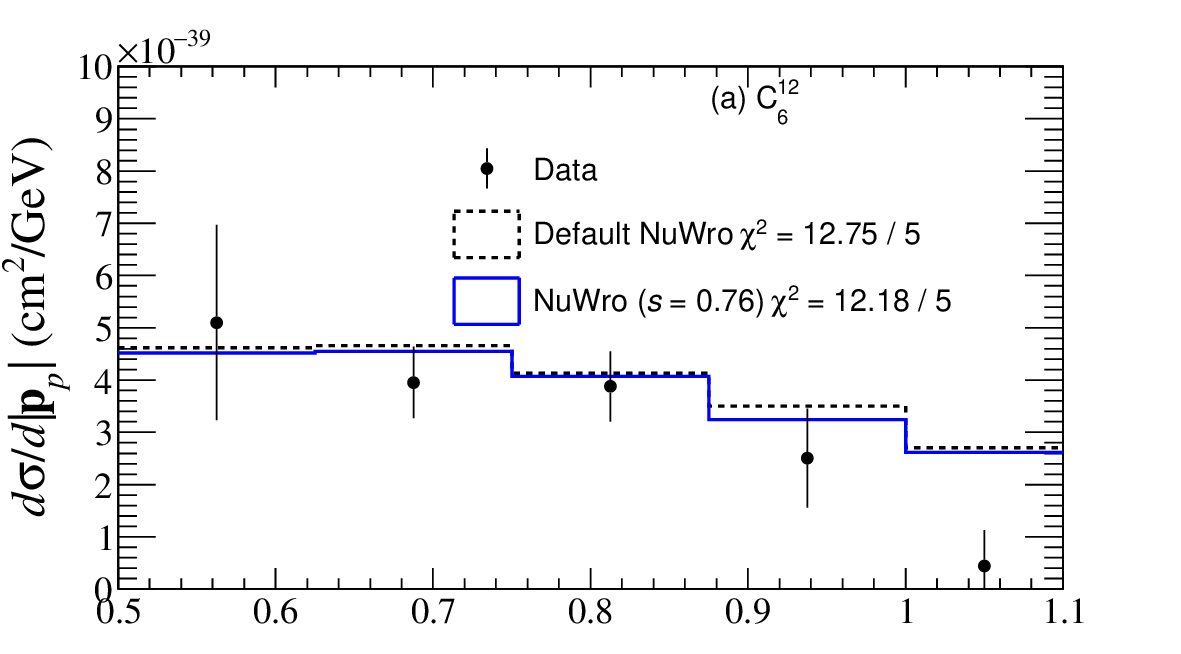}
    \end{minipage}
    \begin{minipage}[b]{\columnwidth}
        \centering 
        \includegraphics[width=1.05\linewidth]{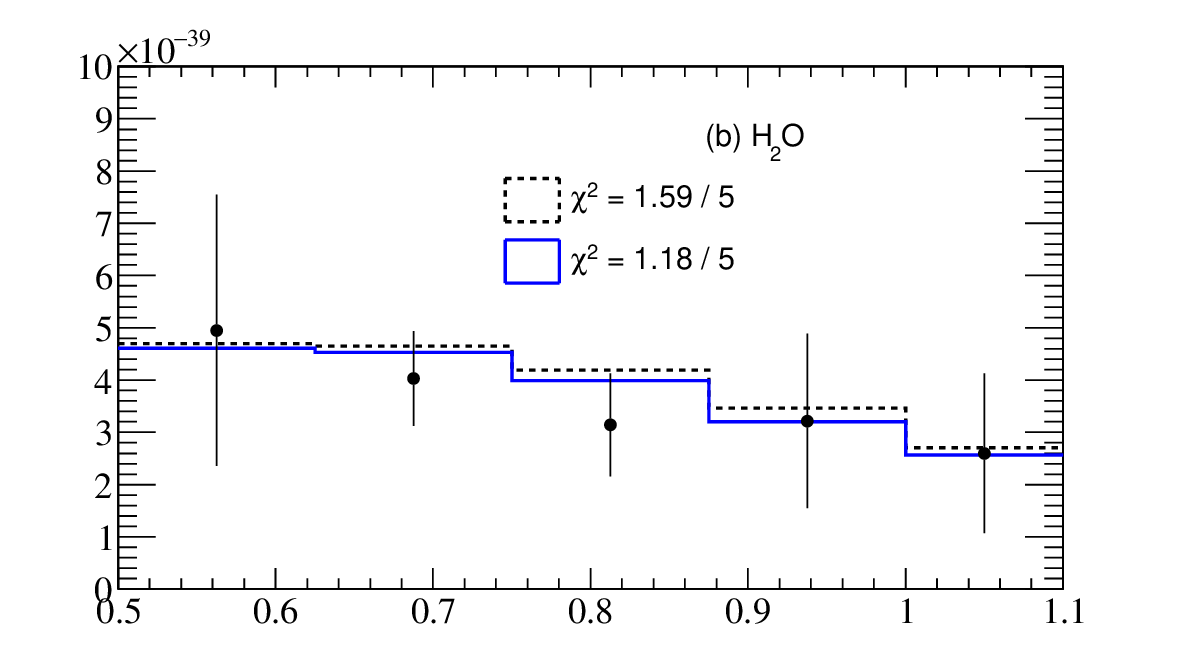}
    \end{minipage}
    \begin{minipage}[b]{\columnwidth}
        \centering 
        \includegraphics[width=1.05\linewidth]{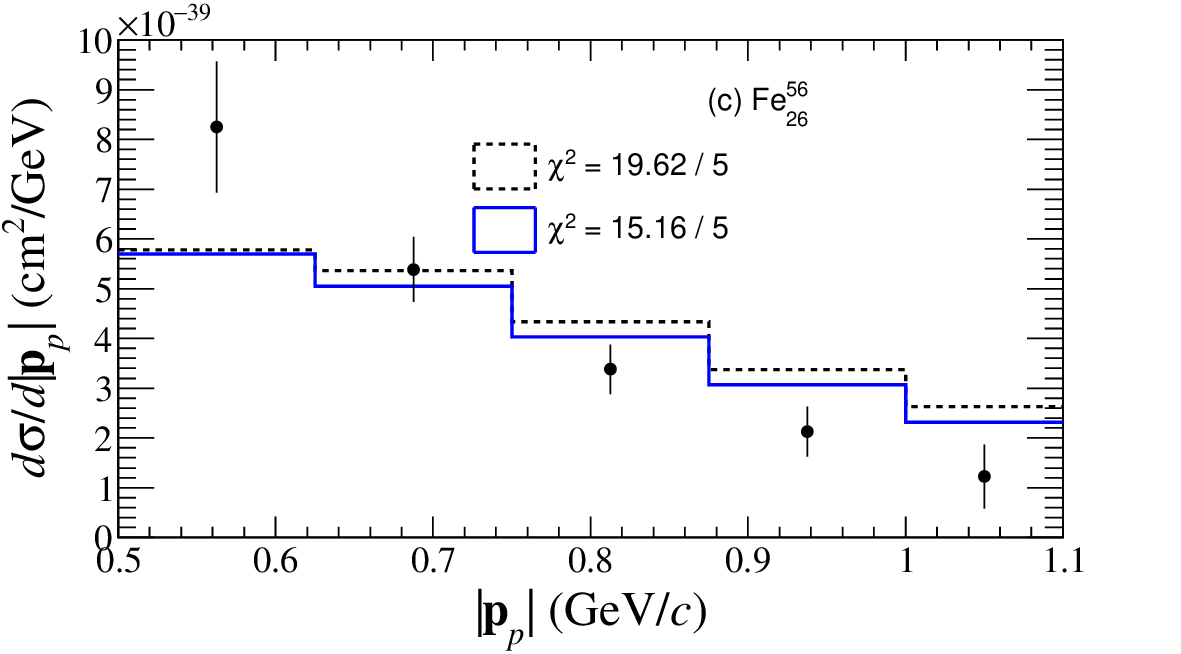}
    \end{minipage}
    \begin{minipage}[b]{\columnwidth}
        \centering 
        \includegraphics[width=1.05\linewidth]{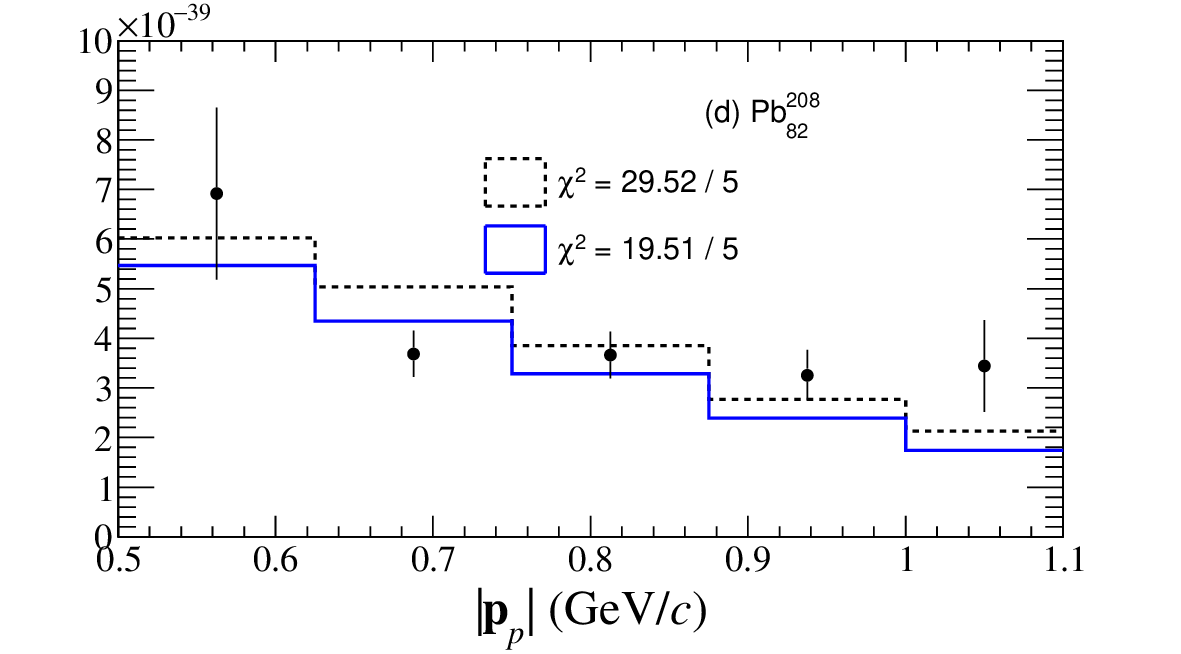}
    \end{minipage}
    \caption{\label{fig:protonmomentum_all_targets_cascde} (Color Online) Same as Fig.~\ref{fig:reconstructedNeutron_all_targets_cascde} but for $|\mathbf{p}_{p}|$}
\end{figure*}

\begin{figure*}[htbp!]
    \begin{minipage}[b]{\columnwidth}
        \centering 
        \includegraphics[width=1.05\linewidth]{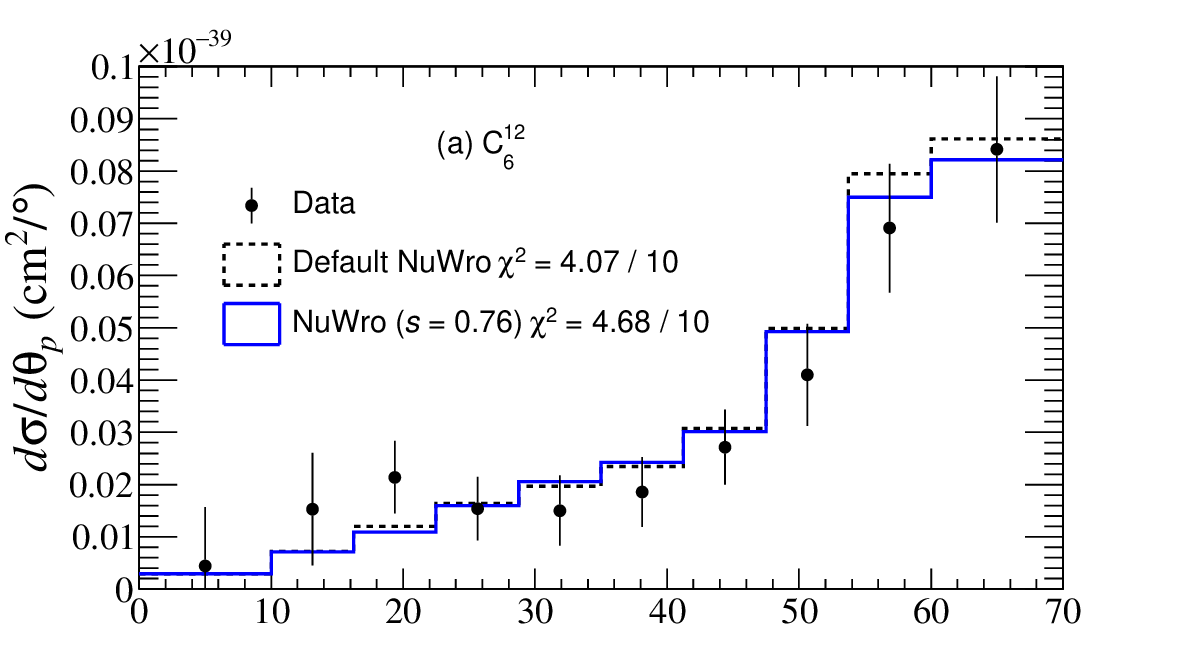}
    \end{minipage}
    \begin{minipage}[b]{\columnwidth}
        \centering 
        \includegraphics[width=1.05\linewidth]{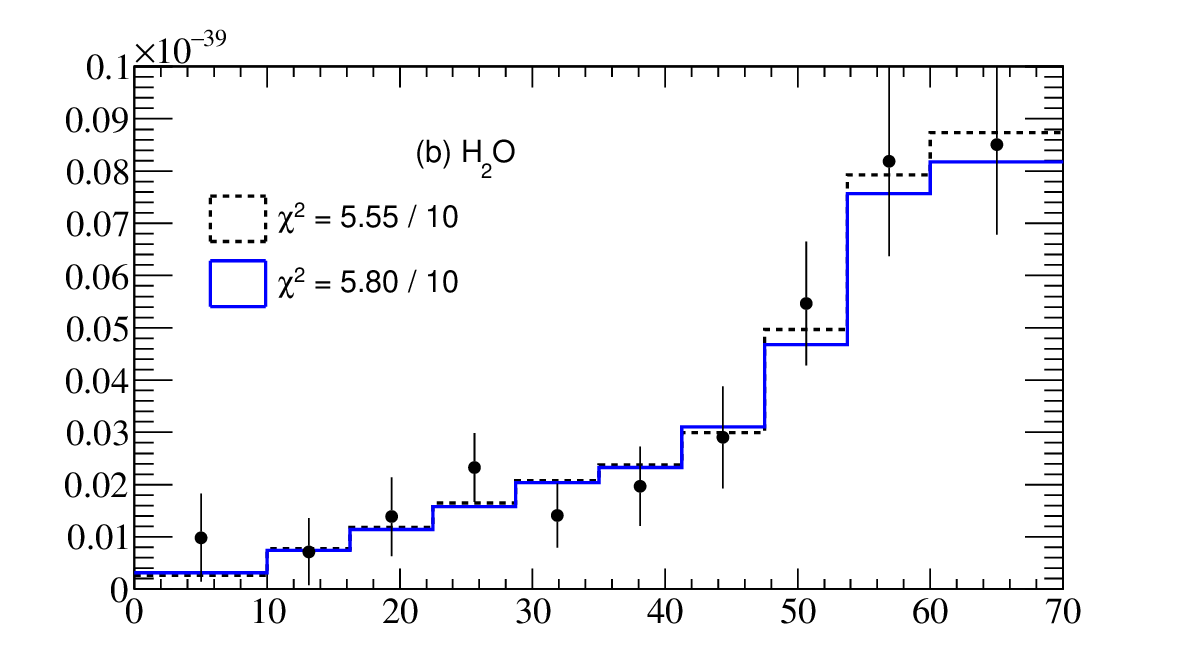}
    \end{minipage}
    \begin{minipage}[b]{\columnwidth}
        \centering 
        \includegraphics[width=1.05\linewidth]{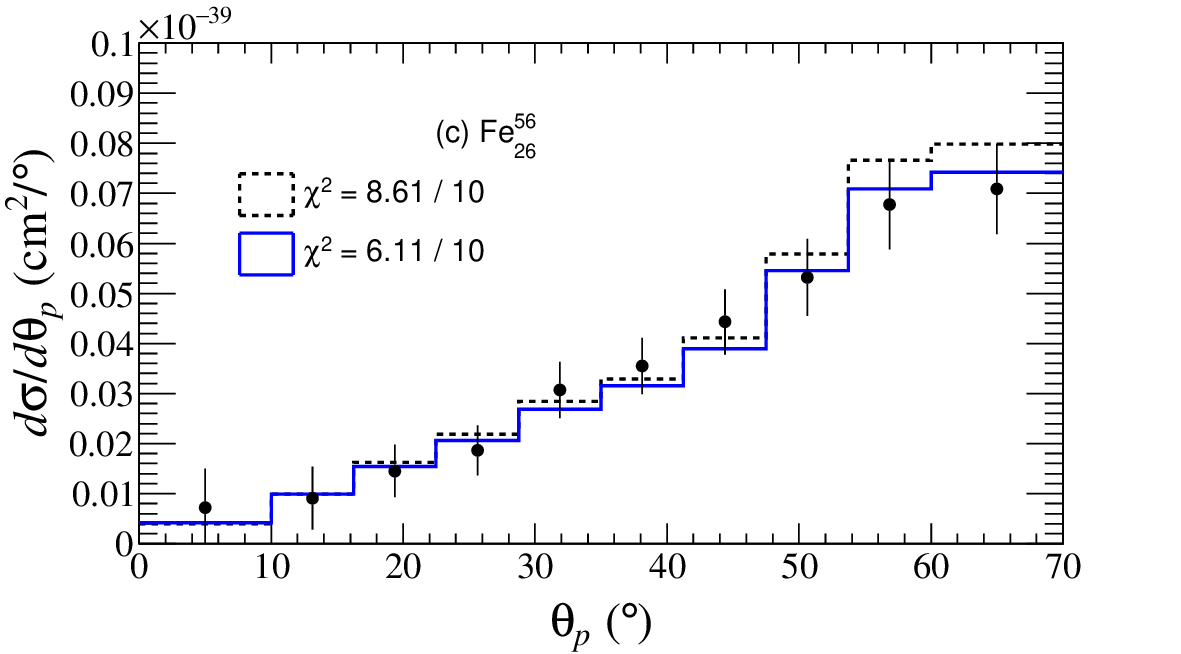}
    \end{minipage}
    \begin{minipage}[b]{\columnwidth}
        \centering 
        \includegraphics[width=1.05\linewidth]{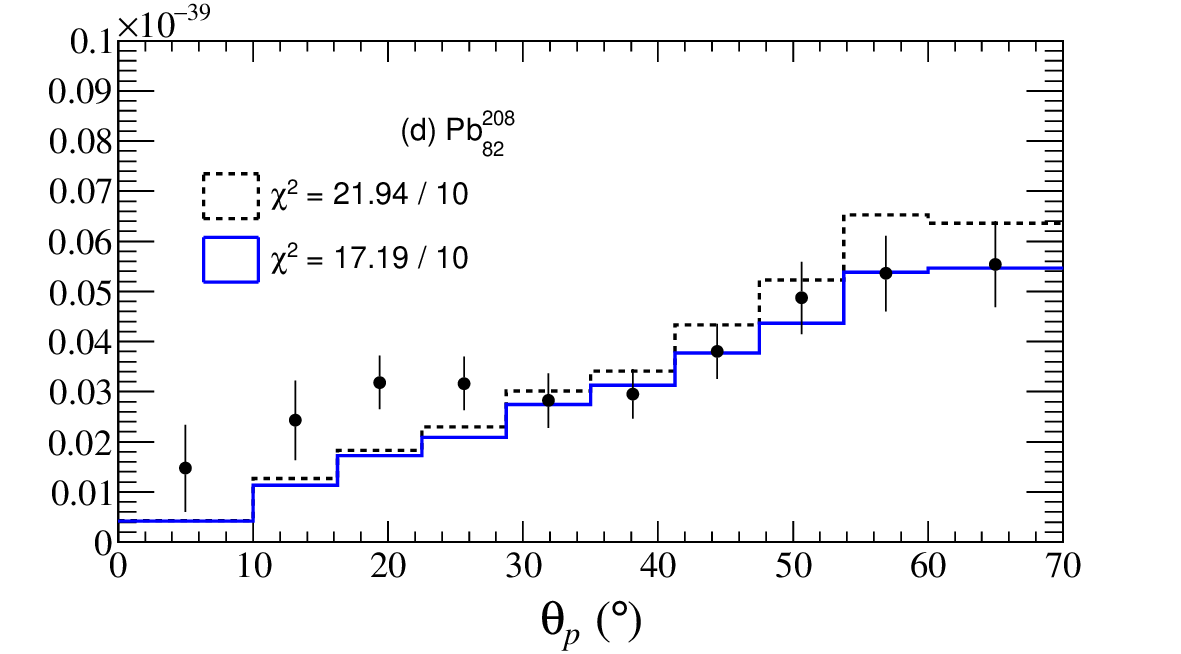}
    \end{minipage}
    \caption{\label{fig:protonangle_all_targets_cascde} (Color Online) Same as Fig.~\ref{fig:reconstructedNeutron_all_targets_cascde} but for $\theta_{p}$}
\end{figure*}

Here we assess \nuwro’s performance for six observables not discussed in Sec.~\ref{subsec:lfg_vs_sf}. Figures~\ref{fig:dalphat_all_targets_cascade}–\ref{fig:protonangle_all_targets_cascde} show the differential cross-section measurements as functions of $\delta\alpha_{T}$, $\delta\mathbf{p}_{T}$, $\delta\mathbf{p}_{T_{y}}$, $\delta\phi_{T}$, $|\mathbf{p}_{p}|$, and $\theta_{p}$, respectively. For almost all observables, the $\chi^{2}$/d.o.f improves after strengthening the \fsi model, with the most significant improvement observed for lead.

Figure~\ref{fig:dpt_all_targets_cascade}(a)–(d) shows \nuwro’s performance for the observable $\delta\mathbf{p}_{T}$. In the absence of nuclear effects, $\delta\mathbf{p}_{T}$ would approximate the initial nucleon momentum and peak around $\sim 0.2$ GeV, indicating the Fermi motion. Figure~\ref{fig:dphit_all_targets_cascade}(a)–(d) presents the differential cross-section measurements as functions of $\delta\phi_{T}$. The persistently poor $\chi^{2}$, even after increasing the \fsi strength, arises mainly from the first bin and reflects limitations of our chosen nuclear model.

Figures~\ref{fig:protonmomentum_all_targets_cascde} and~\ref{fig:protonangle_all_targets_cascde} show \nuwro’s comparison with the \minerva data~\cite{minerva-paper} for the two observables that were not used while fine-tuning $s$ parameter. For both observables, the $\chi^{2}$/d.o.f improves across all nuclear targets when using the best-fit value $s_{0}$ obtained from the six fitted observables.

\bibliography{main.bib}

\end{document}